\documentclass[useAMS,usenatbib]{mn2e}
\usepackage{amsmath}
\usepackage{psfrag} 
\usepackage{natbib}
\usepackage{graphicx}
\usepackage{txfonts}
\usepackage{relsize}
\usepackage{pbox}
\usepackage[usenames,dvipsnames]{color}
\usepackage{bm}
\usepackage{soul}

\newcommand{\Eqref}[1]{Eq.\,(\ref{#1})}
\newcommand{\eqsref}[1]{Eqs.\,(\ref{#1})}
\newcommand{\figref}[1]{Fig.\,\ref{#1}}
\newcommand{\tabref}[1]{Tab.\,\ref{#1}}
\newcommand{\secref}[1]{Sec.\,\ref{#1}}

\newcommand{\inst}[1]{$^{#1}$}

\newcommand{\cross}[1]{}

\newcommand{\trcross}[1]{}

\newcommand{\dvlc}[1]{\textcolor{Black}{#1}}

\newcommand{\vlc}[1]{\textcolor{Black}{#1}}

\definecolor{DarkBrown}{rgb}{.396,.263,.129}
\definecolor{LightBrown}{rgb}{.698,.463,.227}

\newcommand{\half}{\frac{1}{2}}

\newcommand{\tto}{\text{--}}

\newcommand{\AAA}{\mathcal{A}}

\newcommand{\MMM}{\mathcal{M}_{r\phi}}
\newcommand{\MMMterm}{\mathcal{M}_{r\phi}^{\mathrm{term}}}
\newcommand{\tterm}{t_{\mathrm{t}}}

\newcommand{\der}{\mathrm{d}}

\newcommand{\divv}{\nabla \cdot}

\newcommand{\bz}{b_{0z}}

\newcommand{\Ree}{R_{\mathrm{e} }}
\newcommand{\Rm}{R_{\mathrm{m} }}

\newcommand{\tildeRee}{\tilde{R}_{\mathrm{e} }}
\newcommand{\tildeRm}{\tilde{R}_{\mathrm{m} }}

\newcommand{\caz}{c_{\mathrm{A} z}}
\newcommand{\cac}{c_{ \mathrm{Ac}}}

\newcommand{\vek}[1]{\bf{#1}}

\newcommand{\vc}{v_{\mathrm{c}}}
\newcommand{\vp}{v_{\mathrm{p}}}

\newcommand{\tildevc}{\tilde{v}_{\mathrm{c}}}

\newcommand{\bc}{b_{\mathrm{c}}}
\newcommand{\bp}{b_{\mathrm{p}}}

\newcommand{\bfvc}{{\bf v}_{\mathrm{c}}}
\newcommand{\bfvp}{{\bf v}_{\mathrm{p}}}
\newcommand{\tildebc}{\tilde{b}_{\mathrm{c}}}

\newcommand{\bfbc}{{ \bf b}_{\mathrm{c}}}
\newcommand{\bfbp}{{ \bf b}_{\mathrm{p}}}
\newcommand{\gammap}{\gamma_{\mathrm{p}}}

\newcommand{\gammakh}{\gamma_{\mathsmaller{\mathrm{KH}}}}

\newcommand{\kkh}{k_{\mathsmaller{\mathrm{KH}}}}

\newcommand{\lambdamri}{\lambda_{\mathsmaller{\mathrm{MRI}}}}

\newcommand{\gammamri}{\gamma_{\mathsmaller{\mathrm{MRI}}}}

\newcommand{\kmri}{k_{\mathsmaller{\mathrm{MRI}}}}

\newcommand{\vkh}{v_{\mathsmaller{\mathrm{KH}}}}
\newcommand{\dotvkh}{\dot{v}_{\mathsmaller{\mathrm{KH}}}}
\newcommand{\tildevkh}{\tilde{v}_{\mathsmaller{\mathrm{KH}}}}
\newcommand{\bfkh}{{\bf f}_{\mathsmaller{\mathrm{KH}}}}
\newcommand{\bvkh}{ {\bf v}_{\mathsmaller{\mathrm{KH}}}}

\newcommand{\jcop}{J. Comput. Phys.}

\newcommand{\apj}{ApJ}
\newcommand{\mnras}{MNRAS}

\newcommand{\aap}{A{\&}A}

\newcommand{\apjl}{ApJL}

\newcommand{\ie}{i.e.~}

\newcommand{\s}{\textrm{s}}

\newcommand{\ms}{\textrm{ms}}
\newcommand{\km}{\textrm{km}}

\newcommand{\gccm}{\textrm{g cm}^{-3}}

\title[MRI in CCSNe]%
{
On the maximum magnetic field amplification by 
the  magnetorotational instability in
  core-collapse supernovae
}

\author[Rembiasz~et~al.]
{
T.\,Rembiasz\inst{1,2}\thanks{E-mail: tomasz.rembiasz@uv.es},
  J.\,Guilet\inst{1,3},
  M.\,Obergaulinger\inst{2},
  P.\,Cerd\'a-Dur\'an\inst{2}, 
  M.A.\,Aloy\inst{2},
  E.\,M{\"u}ller\inst{1}
  \\
\inst{1} Max-Planck-Institut f{\"u}r Astrophysik, Karl-Schwarzschild-Str.~1, 85748 Garching, Germany \\
\inst{2} Departamento de Astronom\'{\i}a y Astrof\'{\i}sica,  Universidad de Valencia,  C/ Dr.~Moliner 50, 46100 Burjassot, Spain  \\
\inst{3}  Max Planck/Princeton Center for Plasma Physics,  Karl-Schwarzschild-Str.~1, 85748 Garching, Germany \\
}

\begin{document}

\date{Accepted 2016 May 16. Received 2016 April 28; in original form 2016 February 29}

\pagerange{\pageref{firstpage}--\pageref{lastpage}} \pubyear{2016}

\maketitle

\label{firstpage}

\begin{abstract}
  Whether the magnetorotational instability (MRI) can amplify
  initially weak magnetic fields to dynamically relevant strengths in
  core collapse supernovae is still a matter of active scientific
  debate.  \vlc{Recent numerical studies have shown that the first
    phase of MRI growth dominated by channel flows} is terminated by
  parasitic instabilities of the Kelvin-Helmholtz type that disrupt
  MRI channel flows and quench further magnetic field growth.
  However, it remains to be properly assessed by what factor the
  initial magnetic field can be amplified and how it depends on the
  initial field strength and the amplitude of the perturbations.
  Different termination criteria leading to different estimates of
  the amplification factor were proposed within the parasitic
  model. To determine the amplification factor and test which
  criterion is a better predictor of the MRI termination, we perform
  three-dimensional shearing-disc and shearing-box simulations of a
  region close to the surface of a differentially rotating
  proto-neutron star in non-ideal MHD with two different numerical
  codes.  We find that independently of the initial magnetic field
  strength, the MRI channel modes can amplify the magnetic field by,
  at most, a factor of $100$.  Under the conditions found in
  proto-neutron stars a more realistic value for the magnetic field
  amplification is of the order of $10$.  This severely limits the
  role of the MRI channel modes as an agent amplifying the magnetic
  field in proto-neutron stars starting from small seed fields. \vlc{A
    further amplification should therefore rely on other physical
    processes, such as for example an MRI-driven turbulent dynamo.}
\end{abstract}


\section{Introduction}

Originally discovered by \cite{Velikhov__1959__SovPhys__MRI} and
\cite{Chandrasekhar__1960__PNAS__MRI}, the magnetorotational
instability (MRI) was suggested by
\citet[][]{Balbus_Hawley__1991__ApJ__MRI} to be the physical mechanism
driving the redistribution of angular momentum required for the
accretion process in Keplerian discs orbiting compact objects
\citep[see, e.g.\,][for a review]{Balbus_Hawley__1998__RMP__MRI}.

Keplerian discs have a positive radial gradient in angular momentum,
and therefore are linearly (Rayleigh-)stable.  Purely hydrodynamic
perturbations are unlikely to grow to amplitudes at which the
associated stresses can account for efficient angular-momentum
transport.  In the presence of a weak magnetic field, however, a
negative radial gradient in the angular velocity of the disc is
magneto-rotationally unstable, and seed perturbations can grow
exponentially 
on time scales close to the rotational period.  During this phase,
\emph{channel modes} develop. Channel modes are pairs of coherent
radial up- and down-flows stacked vertically and threaded by layers of
magnetic field of alternating radial and azimuthal polarity.  In these
modes, the magnetic tension (Maxwell stress tensor) transports angular
momentum from the inner parts of the disc outwards.

The criterion for the MRI onset can be formulated in a rather simple
manner, even if the thermal stratification (gradients of entropy or
molecular weight) and non-ideal effects (viscosity, resistivity) are
included
\citep{Balbus__1995__ApJ__stratified_MRI,Menou_etal__2004__ApJ__MRI-stability}.
This allows for its application beyond Keplerian discs, in particular
to proto-neutron stars (PNS) resulting from the core-collapse of
rotating massive stars.  Simplified simulations by
\citet{Akiyama_etal__2003__ApJ__MRI_SN} showed that such PNSs possess
regions in which the MRI can grow on shorter time scales than the time
between the bounce and the successful explosion.  This finding, later
confirmed in multi-dimensional models
\citep[e.g.\,][]{Obergaulinger_et_al__2006__AA__MR_collapse_TOV,
  Cerda-Duran_2008,
  Sawai_et_al__2013__apjl__GlobalSimulationsofMagnetorotationalInstabilityintheCollapsedCoreofaMassiveStar,Sawai_Yamada_2015,Mosta_2015},
presents the possibility of generating strong magnetic fields that can
tap the rotational energy of the core, power MHD turbulence
\citep{Masada_et_al__2015}, and become a potentially important
ingredient in rapidly-rotating core collapse supernovae (CCSNe).

How much these systems are affected by the MRI crucially depends on
both its growth rate and on the final amplitude to which the seed
perturbations grow.  We can give an upper limit by assuming that the MRI
ceases to grow once the magnetic field comes close 
to equipartition with the energy of the differential
rotation.  In a CCSN, this would correspond to dynamically important
field strengths of $\sim10^{15} \, \mathrm{G}$ \citep[see
  e.g.\,][]{Meier_etal__1976__ApJ__MHD_SN,
Obergaulinger_et_al__2006__AA__MR_collapse_TOV,
  Cerda-Duran_2008}.

However, this estimate is an upper bound, since it neglects any effect
quenching the MRI before it taps all the available differential
rotation energy.  The physics of the saturation of the MRI growth
remains an active field of research with many studies devoted to
finding the value of the $\alpha$ parameter \citep
{Shakura_Sunyaev__1973__AA__alpha_visco} for the stress tensor.  We
refer, among others, to the works of
\citet{Sano_etal__2004__ApJ__3d-local-MRI-sim-P,Sano_Inutsuka__2001__ApJL__MRI-recurrent-channels,Brandenburg__2005__AN__Turbulence_and_its_parameterization_in_accretion_discs,Fromang_Papaloizou__2007__AA__3d-local-MRI-disc-zero-net-flux_1,Gardiner__2005__MagneticFieldsintheUniverse__Energetics_in_MRI_driven_Turbulence,Knobloch__2005__PhFl__Saturation_of_the_MRI}.

The model of parasitic instabilities by \citet[][GX94
hereafter]{Goodman_Xu}, further studied and developed by
\cite{Latter_et_al}, \citet[][PG09 hereafter]{Pessah_Goodman} and
\citet{Pessah} provides a clear physical picture of the termination
mechanism.  The MRI channel modes are characterised by a shear layer
and a current sheet in the vertical profiles of velocity and magnetic
field, respectively.  Hence, the (laminar) channel flows can be
unstable against \emph{secondary} (or \emph{parasitic})
\emph{instabilities} of Kelvin-Helmholtz (KH) or tearing-mode (TM)
type.
\footnote{\cite{Latter_et_al} classifies the types of parasitic
    modes differently.}
Initially, the role of the parasites is negligible, as they grow
much more slowly than the MRI.  However, since the growth rate of
the secondary instabilities is proportional to the channel mode
amplitude (which grows exponentially with time), it is clear that at
some stage the parasites will grow faster than the MRI channels
(whose growth rate is constant). Roughly at this point, the
parasitic instabilities should disrupt the channel modes and
terminate the MRI growth, marking the transition to the turbulent
saturation phase (PG09).

\citet{Pessah} analytically studied the MRI termination in
resistive-viscous MHD by solving simplified model equations for the
evolution of the parasitic instabilities.  He identified different
parameter-space regimes where, depending on hydrodynamic and magnetic
Reynolds numbers, either the KH instability or the TM is the dominant
(\ie faster developing) secondary instability.  In the regime relevant
for CCSNe, the MRI should be terminated by the KH instability.
  
Some of the predictions of the parasitic model have been confirmed by
\cite{Rembiasz_et_al}, who performed semi-global MRI simulations of a
PNS following \cite{Obergaulinger_et_al_2009}. They found, that given
the CCSNe conditions, the MRI growth is indeed terminated by secondary
parasitic KH instabilities.  Moreover, the properties of these
secondary instabilities were in a good agreement with the parasitic
model.  As expected, the parasitic instabilities developed along the
velocity channels, displaying a very good agreement of the orientation
under which the parasites develop, whereas the wavelength of the
parasitic modes was smaller by a factor of two w.r.t. the theoretical
predictions.

Motivated by this good agreement, we test in this paper further
predictions of the parasitic model, especially those for the maximum
amplification of the Maxwell stress by the MRI channel modes at
termination.  Two different termination criteria were proposed within
the parasitic model (PG09).  Depending on the termination criterion
used in their calculations, \citet{Latter_et_al_2} and \cite{Pessah}
obtained different estimates for the maximum magnetic field
amplification.  So far, these predictions have not been tested with
direct numerical simulations.

In order to properly define the \cross{playground} \vlc{context} in which this paper is
written, 
\vlc{
  we point out that the MRI represents only one way of field
  amplification in CCSNe besides mechanisms such as compression,
  linear winding of poloidal into toroidal field, and hydrodynamic
  instabilities such as convection and the standing accretion shock
  instability (SASI).  The former two do not represent major numerical
  challenges and, hence, their effect on the field strength can be
  understood quite easily, whereas the complexity of the latter
  translates into a significant uncertainty in the factors by which
  the field is amplified with results ranging from a factor of about 5
  \citep{Obergaulinger_2014} to several orders of magnitude
  \citep{Endeve}.  Our work is not concerned with these processes, but
  solely focuses on the MRI, as our numerical methods are not directly
  suited to studying, e.g.\, the large-scale geometry of SASI modes.
  In this context,} 
we should clearly state what we mean by magnetic field amplification,
since there can be more than a single amplification stage.  There is a
primary magnetic field amplification resulting from the exponential
field growth starting from the initial magnetic seed and ending when
coherent MRI channels are disrupted, the field growth saturates and a
turbulent state results.  In the turbulent state, subsequent episodes
of (secondary) MRI amplification may be driven, or a turbulent dynamo
could be formed.  In this work, we only concentrate on the primary
magnetic field amplification, as a further discussion of the MRI
turbulent state is beyond the scope of this paper.

\vlc{As we have discussed in \cite{Rembiasz_et_al}, three dimensional
  (3D) {\em global} simulations to properly assess whether the
  MRI saturates when it begins from {\em realistically} (small)
  magnetic fields of prototypical massive stars are beyond the
  capacities of current supercomputers. Such simulations require a
  prohibitively large spatial numerical resolution. Furthermore,
  a two dimensional (2D) modelling of the saturation process of the MRI
  (in which axial symmetry is assumed for the stellar core) leads to
  qualitatively and quantitatively wrong results. In 2D the dominant
  non-axisymmetric KH modes are suppressed and the MRI is terminated
  by sub-dominant parasitic tearing modes. Consequently, 2D
  simulations overestimate the capability of the MRI to amplify the
  magnetic field \citep{Rembiasz_et_al}. Hence, we are nowadays still
  forced to conduct our studies on this matter employing semi-global
  models as the ones we present here.}

\vlc{The aforementioned limitations of our method make it impossible
  for us to include possibly important effects such as global modes of
  the MRI and the interaction of the MRI with other instabilities.  In
  light of these restrictions, our goal is to find an \emph{upper
    limit} for the field amplification by the MRI channel modes by
  studying the \emph{most favourable} conditions for these channel
  modes to develop.}

\vlc{As we will find that the amplification remains limited even under
  these optimal conditions, we can conclude that channel modes alone
  cannot lead to dynamically relevant field from weak initial
  fields. Other processes are therefore needed, such as for example an
  MRI-driven turbulent dynamo in the turbulent phase that follows
  channel mode disruption.}

In Sec.\ \ref{sec:mri_theory} we describe the initial stage of the MRI
during which channel modes develop.  Next, we discuss its termination
via the parasitic instabilities, examining two different physical
termination criteria.  Finally, we present some estimates for the
maximum magnetic field amplification based on the parasitic model.  In
Sec.\ \ref{sec:numerics}, we describe two different numerical codes
used in our studies and the initial setup of our 3D simulations.  We
present the results of these simulations in Sec.\ \ref{sec:results},
and summarise our findings in Sec.\ \ref{sec:summary}.

\section{MRI growth and termination}
\label{sec:mri_theory}
\subsection{Physical model}
\label{sSec:Physical_Model}

We consider flows that can be described by the equations of
resistive-viscous (non-ideal) magnetohydrodynamics (MHD). In the
presence of an external gravitational potential, $\Phi$, these
equations read
\begin{align}
\label{eq:cont}
\partial_t \rho + \nabla \cdot ( \rho {\vek v} ) &= 0, \\ 
\partial_t( \rho {\vek v }) + 
  \divv \left(\rho {\vek v} \otimes {\vek v} + {\vek T} \right) 
 &= - \rho \nabla \Phi, \\
\partial_{t} e_{\star} +   \divv \left[ 
e_{\star} {\vek v} + {\vek v  \cdot T}
+  \eta \left(  {\vek b} \cdot \nabla {\vek b}  - \mathsmaller{  \frac{1}{2} } \nabla  {\vek  b}^2\right)
  \right]  &= -\rho {\vek v} \cdot \nabla \Phi, \\ 
\partial_t {\vek b} + 
  \divv \left({\vek v}\otimes{\vek b} - {\vek b}\otimes{\vek v} \right) 
 &= \eta \nabla^2 {\vek b}, \\
\divv {\vek b} &= 0,  
\label{eq:divb}
\end{align}
where $\vek v$, $\rho$, $\eta$, and
${\vek b} \equiv {\vek B}/\sqrt{4 \pi}$ are the fluid velocity, the
density, a uniform resistivity, and the redefined magnetic field
${\vek B}$, respectively. The total energy density, $e_{\star}$, is
composed of fluid and magnetic contributions, i.e.\
$e_{\star} = \varepsilon + \frac{1}{2} \rho {\vek v}^2 + \frac{1}{2}
{\vek b} ^ 2$
with the internal energy density $\varepsilon$ and the gas pressure
$p = p(\rho, \varepsilon, \dots)$.  The stress tensor ${\vek T}$ is
given by
\begin{equation}
  {\vek T} = \left[ P + \mathsmaller{\half} {\vek b}^2 
                      + \rho \left( \mathsmaller{\frac{2}{3}}\nu - \xi\right)
                    \divv {\vek v} \right] {\vek I} 
             - {\vek b} \otimes {\vek b} 
             - \rho \nu \left[ \nabla \otimes {\vek v} + (\nabla
               \otimes {\vek v})^T \right], 
\end{equation}
where ${\vek I}$ is the unit tensor, and $\nu$ and $\xi$ are the
kinematic shear and bulk viscosity, respectively.

\subsection{Magnetorotational instability}
\label{sSec:intability_criterion}

We study the MRI in a small portion of the rotating star at a given
distance $r$ from the rotation axis, embedded in a magnetic field. For
convenience, we use cylindrical coordinates $(r,\phi, z)$, hereafter.
We restrict our analysis to locations close to the equatorial plane
($z=0$) and vertical perturbation wavevectors for which the MRI is
known to develop fastest \citep[see,
e.g.\,][]{Balbus_Hawley__1998__RMP__MRI}. In this case, we can consider
a differentially rotating fluid with angular frequency $\Omega$ and
velocity
\begin{equation}
  {\vek v}=  \Omega r {\bm{ \hat{\phi}}},  
\label{eq:v_init}
\end{equation}
threaded by a uniform vertical magnetic field
\begin{equation}
  {\vek b} = \bz {\hat{\vek{z}}} 
\label{eq:b_init}
\end{equation}
in the local perturbation analysis. Here, ${\bm{\hat{\phi}}}$ and
$\hat{\vek{z}}$ are the unit vectors in the direction $\phi$ and $z$,
respectively.
We assume an angular velocity with a radial dependence of the form
\begin{equation}
  \Omega = \Omega_0 \left( \frac{r}{r_0} \right)^{-q},
\label{eq:omega}
\end{equation}
where $\Omega_0$ is the angular velocity at the characteristic radius
$r_0$, and $q$ is the local rotational shear given by
\begin{equation}
  q = - \frac{ \der \ln \Omega}{\der \ln r}. 
\label{eq:q}
\end{equation}

\cite{Balbus_Hawley__1991__ApJ__MRI} investigated the stability of such a system in the limit
of linearised ideal MHD equations, and found
that it is unstable against modes,
usually referred to as \emph{MRI channels},  which grow exponentially with time.
The most unstable MRI mode is characterised by
the vertical wavevector 
\begin{equation}
  \kmri = \sqrt{1 - \frac{(2-q)^2}{4}} \frac{\Omega}{\caz},
\label{eq:kmri}
\end{equation}
where $\caz\equiv b_{0z}/\sqrt{\rho}$ is the Alfv\'en speed in the
vertical direction,
 and grows at a rate
\begin{equation}
  \gammamri = \frac{q}{2} \Omega.
\label{eq:gammamri}
\end{equation}
Its velocity and magnetic field are given by
\begin{align}
  \bfvc(r,\phi,z,t) & = \tildevc e^{\gammamri t} 
               ( {\bf \hat{r}} \cos \phi_v + {\bm{ \hat{\phi}}} \sin \phi_v )
               \sin (\kmri z) , 
\label{eq:channel_v} \\
  \bfbc(r,\phi,z,t) & = \tildebc e^{\gammamri t} 
               ( {\bf \hat{r}}  \cos \phi_b  + {\bm{ \hat{\phi}}} \sin \phi_b )   
               \cos (\kmri z)  
\label{eq:channel_b},  
\end{align}
where the subscript \emph{c} stands for \emph{channel}, $\bf{\hat{r}}$
is the unit vector in $r$ direction, $\tildevc$ and $\tildebc$ are the
initial amplitudes, $\phi_v $ and $\phi_b$ are the angles between the
$r$-axis and the direction of the velocity and magnetic field
channels, respectively. 
To simplify the notation, we define
\begin{align}
  \vc(t)  &= \tildevc  e^{\gammamri t}, \label{eq:vc_t} \\
  \bc(t)  &= \tildebc  e^{\gammamri t},
\end{align}
and for brevity, we will often drop the explicit time dependence,
i.e.\ $\vc = \vc(t)$ and $\bc = \bc(t)$. 
For the fastest-growing mode, the magnetic field and the velocity
amplitudes are related by
\begin{equation}
  v_{\mathrm{c}} = \sqrt{\frac{q}{4-q}} \cac
\label{eq:amplitude_ratio},
\end{equation}
where $\cac \equiv \bc/\sqrt{\rho}$ is the Alfv\'en speed parallel to
the MRI channel, and the channel angles are $\phi_v = \pi/4$ and
$\phi_b= 3\pi/4$.

GX94 showed that MRI modes are exact solutions of the ideal
incompressible MHD equations in the \emph{shearing sheet} (local)
approximation. This approximation consists in transforming the
equations to a frame co-rotating with a fiducial fluid element and
linearising the rotational profile around a radius $r_0$, i.e.\
$\Omega(r) \approx (r-r_0) \partial_r \Omega(r)|_{r_0}$ (for a full
description of this approximation,
cf.\,\citealt{Goldreigh_Lynden-Bell_1965}).  \citet[][PC08
hereafter]{Pessah_Chan} generalised the results of GX94 and showed
that MRI channels (Eqs.\,\ref{eq:channel_v} and \ref{eq:channel_b})
are also exact solutions of the resistive-viscous incompressible MHD
equations in the shearing sheet approximation.  They derived
expressions for the growth rate, the amplitude ratio $\vc / \bc$, and
the channel angles $\phi_v$ and $\phi_b$ of MRI-unstable modes for
arbitrary hydrodynamic and magnetic Reynolds numbers defined as
\begin{align}
  \Ree &= \frac{\caz^2}{\nu \Omega},  
\label{eq:Re}  \\
  \Rm &= \frac{\caz^2}{\eta \Omega}.
 \label{eq:Rm}
\end{align}
Following PC08 and \citet{Rembiasz_et_al}, we use the same definitions
of the Reynolds numbers. We note, however, that
\cite{Guilet_Mueller_2015}, used a different definition of the
Reynolds numbers, namely,
\begin{align}
  \tildeRee &= \frac{L_z^2 \Omega}{\nu},  
\label{eq:tildeRe}  \\
  \tildeRm &= \frac{L_z^2 \Omega}{\eta},
 \label{eq:tildeRm}
\end{align}
where $L_z$ is the vertical size of the computational domain.  We
point out that assuming $L_z = \lambdamri$ and $q = 1.25$, which is
the case in all simulations done with the code \textsc{Snoopy}
presented in this paper (see Sect.\,\ref{sec:Snoopy}), those
differently defined Reynolds numbers are related by
\begin{align}
  \tildeRee &= 46 \Ree,
\\
  \tildeRm &= 46 \Rm.
\end{align}
From now on, we will only use Reynolds numbers defined in
\eqsref{eq:Re} and (\ref{eq:Rm}).

For the Reynolds numbers considered in our studies, i.e.\
$\Ree, \Rm \ge 100$, the characteristics of the fastest-growing MRI
mode (i.e.\ its growth rate, wavelength, and angles) do not change by
more than $\approx 2\%$ with respect to the ideal MHD case (PC08). We
are therefore going to use
\eqsref{eq:kmri}--(\ref{eq:amplitude_ratio}) in our farther analysis.
For a more detailed discussion of non-ideal effects, we refer the
reader to \citet{Rembiasz_et_al}.

\subsection{MRI termination via parasitic instabilities}

GX94 suggested that MRI channels may be unstable against parasitic
instabilities, which could terminate the MRI growth.  They found in
their analytic calculations that in ideal MHD (shear driven) KH modes
can develop on top of MRI channels.  GX94 also suggested that in
resistive MHD, parasitic instabilities of the (current driven) TM type
could develop, too.  Analytical calculations of \cite{Latter_et_al} in
resistive MHD confirmed this hypothesis.  \citet{Pessah} extended
these analytical studies to resistive-viscous MHD.  He identified
regions in parameter space, where depending on the values of the
hydrodynamic and magnetic Reynolds numbers either KH or TM is the
dominant (i.e.\ faster developing) secondary instability that
terminates MRI growth.  In particular, for the conditions prevailing
in CCSNe outside of the neutrinosphere ($\Ree, \Rm \gg 1$) the
exponential growth phase of the MRI should be terminated by KH
instabilities.
This hypothesis was confirmed with direct numerical simulations of
\citet{Rembiasz_et_al}.  In the remaining part of this subsection, we
will first discuss the basic assumptions of the parasitic model then,
we present some estimates which can be derived from it.  A test of
these predictions done with two different numerical codes will be
presented in \secref{sec:results}.

To compute the evolution of the parasitic instabilities,
GX94 considered perturbations in a system with already well developed
MRI channels given by
\begin{align}
  {\vek v} &= -q \Omega_0 (r - r_0) {\bm{ \hat{\phi}}} +    \bfvc(t)    +  \bfvp(r,\phi,z,t)  ,  
\label{channel_full_p1}\\
  {\vek b} &=  b_{0z} {\vek{\hat{z}}} +     \bfbc(t)      +  \bfbp(r,\phi,z,t) ,
\label{channel_full_p2}
\end{align}
where $\bfvp$ and $\bfbp$ are the velocity and the magnetic field of
the parasitic instabilities, respectively. Solving the equations
governing the evolution of the secondary (parasitic) instabilities is
a very challenging task, because MRI channels, which are treated as a
background field for the perturbations, are non-stationary. Hence,
standard analytical techniques cannot be used. To make this task more
tractable for analytic studies, GX94 considered a stage of MRI growth
when the amplitude of the MRI channels is much larger than the initial
weak magnetic field, i.e.\ $ \bc \gg b_{0z}$.  The growth rate of the
secondary instabilities $\gammap$ (which scales $\propto \bc$) is then
much larger than the MRI growth rate, i.e.\ $\gammap \gg
\gammamri$.
Under these conditions the time evolution of the MRI channels, the
Coriolis force, the background shear flow, and the initial background
magnetic field $b_{0z}$ can be neglected because they act on
timescales comparable to $\gammamri^{-1}$.  Hence, instead of
searching for solutions to perturbations according to Eqs.\
(\ref{channel_full_p1}) and (\ref{channel_full_p2}), GX94 considered a
simplified system where the velocity and the magnetic field are given
by
\begin{align}
  {\vek v}(t) &=  \bfvc(t_0)+ \bfvp(r,\phi,z,t) , 
\label{eq:channel_stationary_v} \\
  {\vek b}(t) &= \bfbc(t_0)+ \bfbp(r,\phi,z,t),  
\label{eq:channel_stationary_b}
\end{align}
with $t_0= \mathrm{const.}$ being the time at which the secondary
perturbations are imposed. The same assumptions were also made in the
studies of \citet{Latter_et_al}, PG09 and \citet{Pessah}, even though
these authors extrapolated their results to the regime in which
$\gammap \sim \gammamri$, in which case neglecting the above mentioned
terms is not fully justified. Therefore, the analytical results
obtained by \citet{Latter_et_al} and \citet{Pessah} within the
parasitic model can probably be improved, and direct numerical
simulations should be used to test the former theoretical predictions.

In the ideal MHD limit, the dominant parasitic mode is of the KH type
that develops along the MRI velocity channel (the projection of which
in the horizontal plane, forms an angle $45^{\circ}$ with respect to
the radial direction in the anticlockwise sense) and is characterised
by \citep[e.g.\,][]{Pessah}
\begin{equation}
  \label{eq:kkh}
  \kkh = 0.59 \kmri
\end{equation}
and grows at a rate
\begin{equation}
  \label{eq:gammakh}
  \gammakh = 0.27 \kmri \vc.
\end{equation}
\citet{Rembiasz_et_al} observed in their numerical simulations that
for $\Ree, \Rm \ge 100$ indeed the dominant parasitic modes are of the
KH type and develop along the velocity channels.  However, the
wavelength of the parasitic modes was by a factor of two smaller than
theoretically expected.  As those authors did not investigate in
detail the source of this discrepancy, the question whether it arises
as a result of the numerics or truly from the underlying physics
remains open.

\subsubsection{Termination criteria}
Since the MRI growth rate is constant in time (see \Eqref{eq:gammamri})
and the growth rate of the KH instability grows exponentially with time
(as $\vc \propto \exp(\gammamri t)$), it is clear that at some point
the latter  will exceed the former, i.e.\ $\gammakh > \gammamri$.
Ultimately, the parasitic instabilities will therefore start to drain
more energy from the MRI channels than the MRI can feed into them.
This should eventually lead to the channel disruption (GX94).

PG09 proposed that the MRI is 
terminated at the time, $t_t$, when
\begin{equation}
  \label{eq:term1}
\ \ \ \ \ \ \ \ \ \   \gammakh(t_t) =  \gammamri    \text{ \ \ \ \  \ \ \ \ \ \ \ \ \ \ \ (termination criterion I), }
\end{equation}
According to this criterion, by comparing Eq.\,\eqref{eq:gammakh} with
Eq.\,\eqref{eq:gammamri} (and with the help of Eqs.\ \ref{eq:kmri} and
\ref{eq:amplitude_ratio}), one can find that (independently of the
value of $q$)
\begin{equation}
  \label{eq:b3.8}
  \frac{\bc (\tterm) }{ b_{0z}} = 3.8.
\end{equation}

PG09 also suggested an alternative termination criterion, i.e.\ that
it happens when the amplitudes of the MRI channels and parasitic modes
are comparable, that is
\begin{align}
  \label{eq:term2}
 \vp (\tterm)   &=   \vc  (\tterm)    \text{ \ \ \ or} \\
 \bp   (\tterm)   &= \bc  (\tterm)   \text{ \ \ \ \  \ \ \ \ \ \ \ \ \ \ (termination criterion II), }
\end{align}
where $\vp$ and $\bp$ are the amplitudes of the parasitic
instabilities.  However, they did not investigate this last criterion
in more detail analytically, as in this regime the problem becomes
non-linear. Nevertheless, as pointed out by \citet{Latter_et_al_2},
the termination criterion II seems to be physically much better
justified. Indeed, at the stage when parasites grow at an equal rate
as the MRI channels, but the amplitudes of the parasites are much
smaller (because they have been growing at a lower rate than the MRI),
they can be still treated as small terms that can be
neglected. Therefore, they should be of no physical importance and, in
particular, should not be able to quench the MRI growth.  Using
termination criterion II (and taking into account the influence of the
background shear in an approximate way), \citet{Latter_et_al_2}
estimated that the background shear could increase the termination
amplitude to $\bc (\tterm) /b_{0z} \sim 24\tto 40$ (depending on the
considered amplitude of the parasitic perturbations), which is a
considerably larger value than that estimated using the termination
criterion I (Eq.\,\ref{eq:b3.8}).

Summing up the theoretical estimates, we find, on the one hand, that
termination criterion I most probably underestimates the amplification
of the magnetic field at MRI termination. Nevertheless, it may serve
as a proxy for termination since, as we shall see from our numerical
models, within 2--3 MRI time scales (i.e.\ $\gammamri^{-1}$) after the
growth rates are equal, the MRI channels will be disrupted by the
parasitic instabilities.  On the other hand, in order to use the
termination criterion II, by its own definition, we should not treat
parasitic perturbations as small, linearising the equations in terms
of them, which makes it difficult to provide analytic estimates with a
more physically adequate condition for MRI termination.  This is why
numerical simulations are indispensable to test the predictions of the
parasitic model.

\subsubsection{Some estimates}

We define the absolute value of the volume-averaged Maxwell stress
component as
\begin{equation}
  \MMM \equiv \frac{ \left| \int b_r b_{\phi} \ \mathrm{d} V \right|}{V} ,
\label{eq:MMM}
\end{equation}
where $V$ is the volume of the computational domain, and
the amplification factor as
\begin{equation}
  \label{eq:AAA}
  \AAA \equiv  \frac{ \sqrt{ \MMMterm}  }{b_{0z}},
\end{equation}
where $\MMMterm \equiv \MMM(\tterm)$.  Assuming that at termination,
the MRI channels are still given by \Eqref{eq:channel_b}
(i.e.~undistorted) and ignoring the contribution of the parasitic
instabilities, the Maxwell stress is
\begin{equation}
  \label{eq:MMM_term}
\MMMterm = \frac{\bc^2(\tterm)}{4}.
\end{equation}
Hence, from the estimate done by 
PG09 using the termination criterion I (Eq.~\ref{eq:b3.8}),
we obtain
\begin{equation}
  \label{eq:a1.9}
  \AAA  =  1.9.
\end{equation}

Next, we want to make some estimates of the MRI termination amplitude
within the parasitic model using the second termination criterion
(Eq.~\ref{eq:term2}).  Note that we begin our analysis from the
equations obtained by \cite{Pessah} under some simplifying assumptions
and that we will further introduce additional
simplifications. Therefore, our estimates must be tested with the
simulation results presented in \secref{sec:results}.

In order to make use of the second termination criterion, we need to
first find the time evolution of the parasitic KH modes.  For this
purpose, we assume that during their whole evolution, i.e.\ from their
onset until the MRI termination, they can be factored in the following
terms
\begin{equation}
  \label{eq:4}
  \bvkh(r,\theta,z,t) = \vkh(t) \bfkh(r,\theta,z),
\end{equation}
where $\vkh(t)$ is the instability amplitude and includes the time
evolution of the perturbation, and $\bfkh$ is a normalised function
(i.e.\ $\mathrm{max}(\bfkh)=1$) that does not depend on time
\citep[its explicit form can be found in][]{Pessah}.  Note that the
factorisation assumed in Eq.\,(\ref{eq:4}) may fail because, among
other reasons, (i) we neglect effects proportional to $\Omega$ (the
shear and the Coriolis force), and (ii) because shortly before the MRI
termination, when $\vkh \approx \vc$, the KH instability will enter
the non-linear phase of its evolution.  In this phase the KH
perturbations eventually become comparable to the background shear,
and the growth of the instability is reduced before its final
termination \citep[cf.][]{Keppens_et_al,Obergaulinger_et_al_KH}.

We can find the time evolution of $\vkh(t)$ from the definition of the
KH instability growth rate, i.e.
\begin{equation}
  \label{eq:gammakh_t}
  \gammakh(t)\equiv  \frac{  \dotvkh(t)  }{ \vkh(t) },
\end{equation}
provided that  $  \gammakh(t)$ is known.
The resulting differential equation
can be integrated analytically if we assume
$\gammakh(t) = \sigma \kmri \vc(t)$, where $\sigma = \mathrm{const.}$%
\footnote{For $\sigma = 0.27$, we recover \Eqref{eq:gammakh},
however for the time being we want to use a bit more general expression.},
that is \citep[see][for a similar calculation on a different primary
instability]{latter16}
\begin{equation}
  \label{eq:6}
  \vkh(t) = \vkh(t_0) \exp \left[  \frac{ \sigma \kmri \vc(t_0)} {\gammamri}  \left( e^{\gammamri t} - e^{\gammamri t_0} \right) \right],
\end{equation}
where $t_0$ is the time at which the KH perturbations begin to grow.
Assuming $t_0 = 0$, we finally obtain
\begin{equation}
  \label{eq:vkh_time}
  \vkh(t) = \tildevkh \exp \left[  \frac{ \sigma \kmri      \tildevc }{\gammamri}  \left( e^{\gammamri t} - 1 \right) \right],
\end{equation}
where $\tildevkh$ is the initial KH amplitude.

Now, we can determine the MRI termination amplitude from the condition
\begin{equation}
  \label{eq:term_2_s}
\frac{  \vkh(\tterm) }{ \vc(\tterm) } = s,
\end{equation}
where for $s = 1$, we recover the termination criterion II exactly,
but we allow this parameter to slightly vary, as the choice $s = 1$ is
somewhat arbitrary.  With the help of \eqsref{eq:vc_t} and
(\ref{eq:vkh_time}), we can rewrite the above condition as
\begin{equation}
  \label{eq:term2_step}
\tildevc  e^{  \gammamri \tterm }=  \frac{ \gammamri }{\sigma \kmri  }  \left[  \ln \left(  \frac{ s  \tildevc}{ \tildevkh} \right)+  \gammamri \tterm \right] + \tildevc
\end{equation}
where the term on the LHS of Eq.\,(\ref{eq:term2_step}) is equal to
the definition of the velocity amplitude at termination $\vc(\tterm)$
(see, Eq.\,\ref{eq:vc_t}).

We note that Eq.\,(\ref{eq:term2_step}) should be treated as an
equation for $\tterm$. There is a trivial solution, $\tterm = 0$, for
$\tildevkh = s \tildevc $. However, these are not physically plausible
initial conditions.  For the MRI channels to be destroyed by the KH
instability, they should exist in the first place.  So even if we
start with completely random initial perturbations (which is
physically plausible), we first expect the MRI channels to develop,
implying that $\tterm>0$. Then, after a sufficiently developed
velocity shear sets in between the channels, the parts of the initial
perturbations which do not promote the growth of the MRI will seed the
growth of the KH instabilities. In practical terms, this means that we
can safely assume that $\tildevkh < s \tildevc $ ($s\lesssim 1$). From
\eqsref{eq:amplitude_ratio} and (\ref{eq:MMM_term}), we find
\begin{equation}
  \label{eq:vc_term_step}
  \vc (\tterm)  
 = \sqrt{ \frac{4  q }{4 - q} }  \AAA \caz .
\end{equation}
Comparing \Eqref{eq:vc_term_step}  with \Eqref{eq:term2_step}, we  obtain
\begin{equation}
  \label{eq:aaa_time}
  \AAA =  \frac{1}{2 \sigma}   \left[  \ln \left(  \frac{ s  \tildevc}{ \tildevkh} \right)+  \gammamri \tterm \right] +  \sqrt{ \frac{4 - q}{4  q }  }  \frac{\tildevc}{\caz},
\end{equation}
where we also used \eqsref{eq:kmri} and (\ref{eq:gammamri}).  From
\eqsref{eq:vc_t}, evaluated at $t=\tterm$, and
(\ref{eq:vc_term_step}), we find
\begin{equation}
  \label{eq:3}
  \gammamri t = \ln \left( \sqrt{ \frac{ 4q }{ 4 - q}}  \AAA  \frac{ \caz }{ \tildevc} \right)
\end{equation}
that we substitute into \Eqref{eq:aaa_time}, to finally obtain
\begin{equation}
  \label{eq:AAA_ln}
  \AAA - \frac{1}{2 \sigma} \ln \AAA =  \frac{1}{2 \sigma}   \left[  \ln \left( \frac{  \caz }{ \tildevkh } \right) +
    \ln \left(s \sqrt{ \frac{ 4q }{ 4 - q}} \right) \right] +  \sqrt{ \frac{4 - q}{4  q }  }  \frac{\tildevc}{\caz}.
\end{equation}
Since, typically $\tildevc / \caz \ll 1$ and $ \caz /\tildevkh \gg 1$,
the amplification factor should be almost independent of the initial
MRI channel amplitude and depend logarithmically on the ratio of the
initial velocity amplitude of the parasitic instabilities to the
amplitude of the background Alfv\'en speed (or, equivalently, the
initial magnetic field strength).  For $\sigma = 0.27$ (the value
calculated by PG09), $s=1$ and $ \caz/\tildevkh = 1000$ (a typical
value used in our simulations), we obtain
\begin{equation}
  \label{eq:a19}
  \AAA \simeq 19.
\end{equation}

Finally, we discuss the results of \citet{Latter_et_al_2} who gave an
approximate description of the influence of the background shear on
the development of the parasitic instabilities. They pointed out that
due to shear, the radial wavenumber of non-axisymmetric parasitic
modes increases linearly with time while the azimuthal wavenumber
remains unchanged.  As a consequence, during most of its evolution,
the parasitic mode is expected to grow at a rate smaller than that
predicted by GX94, since its wavelength and orientation w.r.t.\ the
velocity channel is not optimal.  Moreover, the parasitic mode can
only grow for a limited time $\tau$ before the shear prevents its
further growth completely.  \citet{Latter_et_al_2} found that the
dominant mode would have at most a time of
\begin{equation}
  \label{eq:tau}
\tau = \frac{2.18}{q \Omega }
\end{equation}
for its development. %
Furthermore, they roughly estimated that the growth rate of the mode
is reduced by a factor of two as a result of the shear, i.e.
\begin{equation}
  \label{eq:8}
  \gammakh(t) = \frac{\sigma}{2} \kmri \vc(t),
\end{equation}
and calculated the MRI termination amplitude, from the condition\,
(\ref{eq:term_2_s}) assuming that parasitic perturbations are
introduced at $t = t_0 \neq 0$ and that $\tau=\tterm-t_0$.  If within
the time interval $\tau$, condition (\ref{eq:term_2_s}) is not met,
the MRI will not be terminated.  Hence, these authors determined the
minimum amplitude of the velocity perturbations $\vc(t_0)$ for which
the parasitic modes can catch up with the MRI channels after the
aforementioned time interval $\tau$ has passed. When calculating the
growth rate of the parasitic instabilities, \citet{Latter_et_al_2}
further neglected the fact that within the time interval $\tau$ the
parasitic growth rate increases as the MRI channel amplitude increases
with time. They thus obtain
\begin{equation}
  \label{eq:lat_const}
  \vkh(t)= \vkh(t_0) \exp\left[ \frac{ \sigma }{2}\kmri \vc(t_0) (t - t_0)\right].
\end{equation}
By plugging \Eqref{eq:lat_const} to
\Eqref{eq:term_2_s}, we have
\begin{equation}
  \label{eq:7}
  \frac{\vkh(t_0)}{\vc(t_0)} \exp \left [\tau \left( \frac{ \sigma}{2} \kmri \vc(t_0) -    \gammamri\right) \right] = s,
\end{equation}
from which we can calculate (using also Eqs.\,\ref{eq:kmri},
\ref{eq:gammamri} and \ref{eq:amplitude_ratio})
\begin{equation}
  \label{eq:AAA_Lat_0}
  \AAA = \frac{1}{\sigma}\left[ 0.92 \ln\left( \frac{ s \vc(t_0)}{
        \vkh(t_0) } \right) + 1 \right].
\end{equation}
In order to make a further progress with this equation, we have to
calculate $\vkh(t_0)$, ($\vc(t_0) = \tildevc e^{\gammamri t_0}$ from
Eq.\ \ref{eq:vc_t}).  By assuming that growth of the parasitic
instabilities from time $t = 0$ to $t = t_0$ is negligible, i.e.\
$\vkh(t_0) \approx \vkh(0) = \tildevkh$, which is well justified in
the light of our simulations done with Snoopy (see
\figref{sec:Snoopy_sim} of model \#S19 discussed in
\secref{fig:parasitic_Snoopy}), we finally obtain
\begin{equation}
  \label{eq:AAA_Lat}
  \AAA - \frac{0.92}{\sigma}\ln\AAA= \frac{0.92}{\sigma}\left[  \ln \left( \frac{  \caz }{ \tildevkh } \right) +
    \ln \left(s \sqrt{ \frac{ 4q }{ 4 - q}} \right) \right] + \frac{1}{\sigma}.
\end{equation}
This equation has a very similar form to Eq.~\ref{eq:AAA_ln}, but
predicts larger amplification factors by a factor $\approx2$.  For
typical values used in our simulations (see the discussion above Eq.\
\ref{eq:a19}), we obtain
\begin{equation}
  \label{eq:a41}
  \AAA \simeq 41.
\end{equation}
Note, however, that as in \citet{Latter_et_al_2}, it was assumed that
$\bc(\tterm) \approx \bc(t_0)$, i.e.\ the growth of the magnetic field
in the time interval $\tau$ was neglected.  Taking this effect into
account would increase the estimate of $\AAA$ by another factor of
$\exp{(\gammamri \tau)}\simeq 3$, i.e.\
\begin{equation}
  \label{eq:a123}
  \AAA \simeq 123.
\end{equation}

\section{Numerical setup}
\label{sec:numerics}

To test the predictions of the parasitic model, we perform simulations
using two different numerical codes, a finite-volume code,
\textsc{Aenus}, and a pseudo-spectral code, \textsc{Snoopy}, that we
describe briefly in the following subsections. The advantages and
disadvantages of the simulations performed with these codes will be
discussed in \secref{sec:results}.

\subsection{Numerical codes}

\subsubsection{Aenus}
\label{sec:aenus}

We use the three-dimensional Eulerian MHD code \textsc{Aenus}
\citep{Obergaulinger__2008__PhD__RMHD} to solve the MHD equations
(\ref{eq:cont})--(\ref{eq:divb}).  The code is based on a
flux-conservative, finite-volume formulation of the MHD equations and
the constrained-transport scheme to maintain a divergence-free
magnetic field \citep{Evans_Hawley__1998__ApJ__CTM}.  The code is
based on high-resolution shock-capturing methods
\citep[e.g.\,][]{LeVeque_Book_1992__Conservation_Laws}. It implements
various optional high-order reconstruction algorithms, including a
total-variation-diminishing piecewise-linear (TVD-PL) reconstruction
of second-order accuracy, a third-, \mbox{fifth-,} seventh- and
ninth-order monotonicity-preserving (MP3, MP5, MP7 and MP9,
respectively) scheme \citep{Suresh_Huynh__1997__JCP__MP-schemes}, a
fourth-order, weighted, essentially non-oscillatory (WENO4) scheme
\citep{Levy_etal__2002__SIAM_JSciC__WENO4}, and approximate Riemann
solvers based on the multi-stage (MUSTA) method
\citep{Toro_Titarev__2006__JCP__MUSTA}.
We add terms including viscosity and resistivity to the flux terms in
the Euler equations and to the electric field in the MHD induction
equation.  We treat these terms similarly to the fluxes and electric
fields of ideal MHD.  The derivatives of velocity and magnetic field
appearing in the viscous fluxes and resistive electric field,
respectively, are computed from reconstructed states obtained by same
high-order reconstruction methods as for the terms of ideal MHD.  The
explicit time integration can be done with Runge-Kutta schemes of
first, second, third, and fourth order (RK1, RK2, RK3, and RK4),
respectively.

We performed the simulations reported here with the MP9 scheme, a
MUSTA solver based on the HLLD Riemann solver, and an RK3 time
integrator \citep[]{Harten_JCP_1983__HR_schemes,HLLD}. See
\citet{Rembiasz_et_al} for a justification of this choice.

\subsubsection{Snoopy}
\label{sec:Snoopy}

We use the pseudo-spectral code \textsc{Snoopy}
\citep{lesur05,lesur07} to solve the MHD equations
(\ref{eq:cont})--(\ref{eq:divb}) in the shearing box and
incompressible approximations. The incompressible approximation holds
if both the flow and Alfv\'en velocities are much less than the sound
speed. It further assumes a uniform background density $\rho_0$ and
entropy, \ie one neglects the radial density and entropy gradients.
The incompressible approximation can be considered a special case of
the Boussinesq approximation (when the entropy gradient vanishes),
whose validity in the CCSN conditions has been extensively discussed
in \citet{Guilet_Mueller_2015}. \textsc{Snoopy} solves the 3D shearing
box equations using a spectral Fourier method, where the shear is
handled through the use of a shearing wave decomposition (with time
varying radial wavevector) and a periodic remap procedure. Nonlinear
terms are computed with a pseudo-spectral method using the 2/3
dealiasing rule. The time integration is performed using an implicit
procedure for the diffusive terms, while other terms use an explicit
3rd order Runge Kutta scheme. \textsc{Snoopy} is parallelised using
both MPI and OpenMP techniques. It has been extensively used in the
past to study the MRI in the context of accretion discs
\citep{lesur07,longaretti10,rempel10,lesur11,meheut15,walker15} and
PNSs \citep{Guilet_Mueller_2015}.

\subsubsection{Initial conditions}

In the simulations performed with \textsc{Aenus}, following
\citet{Rembiasz_et_al} and \citet{Obergaulinger_et_al_2009}, we use
equilibrium initial conditions based on the final stages of
post-bounce cores from
\cite{Obergaulinger_Aloy_Mueller__2006__AA__MR_collapse}, in which
(several tens of milliseconds after core bounce) the shock wave has
reached distances of a few hundred kilometres and the post-shock
region exhibits a series of damped oscillations as the PNS relaxes
into a nearly hydrostatic configuration.
The rotational profile (given by Eq.\,\ref{eq:omega} with $\Omega_0 =
767\, \s^{-1}$, and $q = 1.25$) that we used in our simulations, is
similar to the one employed in the global MRI simulations of
\citet{Obergaulinger_Aloy_Mueller__2006__AA__MR_collapse}.  Because
the resulting centrifugal force is insufficient to balance gravity,
the gas is kept in (an initial hydrodynamic) equilibrium by an
additional pressure gradient, so that
\begin{equation}
  \rho \partial_r \Phi  -\partial_r P + r \rho \Omega^2 = 0.
\label{eq:equilibrium}
\end{equation}
The initial distributions of angular velocity, density, and
gravitational potential were obtained by rescaling those used by
\citet[][Fig.\,1]{Rembiasz_et_al} to match the conditions encountered
at the PNS surface (see Appendix \ref{app:conversion} for more
details).

At this location, neutrinos do not have a strong impact on the
dynamics \citep{Guilet_et_al_2015} and we therefore expect a very
small viscosity. In all \textsc{Aenus} simulations (see
\tabref{tab:aenus}), we set the shear and bulk viscosity to $\nu = \xi
=0\, \mathrm{cm}^2\,\mathrm{s}^{-1}$, and the value of resistivity was
chosen so that the magnetic Reynolds number $\Rm = 100$.
Therefore, if we consider, e.g.\, an initial magnetic field
strength $b_{0z} = 1.22 \times 10^{13}\, \mathrm{G}$, the previous
value of $\Rm$ corresponds to $\eta = 7.48 \times 10^{8}\,
\mathrm{cm}^2\,\mathrm{s^{-1}} $.  The typical simulation domain was
set to $L_r \times L_{\phi} \times L_z = 2\,\km\times 2\,\km\times
\lambdamri$, except for the models \#A7, \#A8a,b,c, \#A11, in which
$L_r \times L_{\phi} \times L_z = 2\,\km\times 8\,\km\times 3
\lambdamri$.  Note however, that both $\Rm$ and $\lambdamri$ are not
constant in the computational domain but vary by some $\approx 20\%$
\citep[see][for details]{Rembiasz_et_al}.

We assume periodic boundary conditions in the directions $\phi$ and
$z$.  In the radial direction, we use shearing disc boundary
conditions
\citep{Klahr_Bodenheimer__2003__ApJ__Global-baroclinic-inst-disc},
i.e.\,we apply periodic boundary conditions to the deviation of
several variables from their initial state. For instance, applied to
the fluid density we have that such deviation is
\begin{equation}
  \delta \rho(r,t) \equiv \rho(r,t) - \rho(r,0),
\label{eq:deltarho}
\end{equation}
and we enforce shear-periodicity of the variables $\delta \rho(r,t)$
\citep[see][for a more detailed justification of this
choice]{Rembiasz_et_al}. We apply these boundary conditions to angular
velocity, density, momentum, and entropy.  Because the initial
magnetic field is homogeneous in all our simulations, we use periodic
boundary conditions for this quantity too.

In all \textsc{Snoopy} simulations (see \tabref{tab:Snoopy}), we use
analogous initial conditions with the exceptions that simulations are
done in the frame corotating with the fluid and the rotational profile
is linearised (shearing box approximation).  Moreover, the background
density is uniform and set to $\rho_0 = 2.47 \times 10^{12}\,\gccm$
(which corresponds to the central value in \textsc{Aenus}
simulations), and the shear viscosity and resistivity are chosen so
that $\Ree = \Rm = 100$.

In the simulations done with \textsc{Aenus}, following
\citet{Rembiasz_et_al}, to trigger the MRI we impose an initial
velocity perturbation on the background velocity profile (defined by
Eq.\,\ref{eq:v_init}) of the form
\begin{align}
  {\vek v_1}= \Omega r \big[ & \{ 
               {\delta}_{r} \mathfrak{R}_{r}(r,\phi,z) +
               \epsilon \sin(k_z z) \} {\vek{\hat{r}}} + 
\nonumber \\
             & 
               \delta \mathfrak{R}_{\phi}(r,\phi,z) 
                      {\vek{\hat{\bm{\phi}}}} +  
               \delta \mathfrak{R}_{z}(r,\phi,z) {\vek{\hat{z}}} \big],  
\label{eq:v_with_sin}
\end{align}
where $\mathfrak{R}_{r}(r,\phi,z)$, $\mathfrak{R}_{\phi}(r,\phi,z)$,
and $\mathfrak{R}_{z}(r,\phi,z)$ are random numbers in the range
$[-1,1]$, $\delta$ and $\delta_r$ are the perturbation amplitudes,
$k_z$ is the vertical perturbation wavenumber, and $\epsilon$ is the
amplitude of the sinusoidal perturbation in the $z$-direction.  
We choose $\delta_r = 0.1 \delta$. Typically, $\epsilon$
and $\delta$ are of the order of $10^{-5}$. Their exact values in each simulation
can be found in Tab.\,\ref{tab:aenus}.

\vlc{The random perturbations added to the channel modes in our
  simulations are rather small compared with the actual perturbations
  (of the order of one) expected in the collapsed core of a massive
  star. Larger perturbations result in shorter periods of magnetic
  field growth by the action of MRI channels.  Thus, we expect that
  our numerical results set an upper bound for the field
  amplification by MRI channels in the collapsed core of massive
  stars.}

In all simulations we set $k_z = \kmri$, with the only exception being
model \#A9b in which $k_z = 3 \kmri$.  Even though we initialise the
channel only in one velocity component instead of both radial and
azimuthal components of velocity and magnetic field, the other
components quickly grow and form a fully developed channel mode.

In \textsc{Snoopy} simulations, we use five different prescriptions
for the initial perturbations.  In simulations \#S1a--\#S18 from
\tabref{tab:Snoopy}, to the background velocity,
$ {\vek v_0} = -q \Omega_0 (r - r_0) {\bm{ \hat{\phi}}}$, we added
``one component of an MRI channel'', i.e.
\begin{equation}
{\vek v_{\epsilon}}=   \epsilon \Omega_0 r_0     {\bm{ \hat{r}}}   \sin ( \kmri z)
\end{equation}
and random perturbations of the form
\begin{equation}
\tilde{\vek v}_{\vek 1} =   \delta \Omega_0 r_0  \sum_{\beta}    {\bm{ \hat{\beta}}}  \sum_{l, m, n} 
                      \mathfrak{R}^{lmn}_{\beta} \sin (k_{l} r +   k_{m} \phi +
                                     k_{n} z + \chi^{lmn}_{\beta})
        \label{eq:Snoopy_random}
\end{equation}
where $\beta \in \{ r,\phi,z \}$, $k_{\beta } = 2\pi \,l / L_{\beta}$,
$\mathfrak{R}_{lmn}$ and $\chi_{lmn}$ are random amplitude and phase
of a given mode. The sum is restricted to Fourier modes with
wavelength longer than $0.1 \, \km$. The amplitudes
$\mathfrak{R}^{lmn}_{\beta} $ are random numbers between $0$ and
$\delta \Omega_0 r_0/\sqrt{n_{\rm pert}}$, where $n_{\rm pert}$ is the
number of excited modes. The solenoidal nature of the velocity field
in the incompressible approximation is enforced \textit{a posteriori}
by subtracting the divergence part of the field. Random magnetic field
perturbations ${\vek b_{1}}$ are constructed in an analogous way, the
Alfv\'en velocity replacing the velocity in the equations.
Hence the final form of the initial velocity and magnetic field is
\begin{align}
  {\vek v} &=    -q \Omega_0 (r - r_0) {\bm{ \hat{\phi}}}        +
  {\vek v_\epsilon}    +     \tilde{\vek v}_{\vek 1} ,
\label{eq:Snoopy_v}
\\
  {\vek b} &=  b_{0z} {\vek{\hat{z}}}     +    {\vek b_1}.
\label{eq:Snoopy_b}
\end{align}
The values of $\delta$ and $\epsilon$ in each simulation can be found
in \tabref{tab:Snoopy}. The perturbations in simulations \#S19 follow
from the same procedure except that all wavelengths down to the grid
scale are perturbed.  In simulations \#SA15a--e from
\tabref{tab:Snoopy} the random velocity perturbations
$\tilde{\vek v}_{\vek 1}$ are replaced by cell-by-cell random values
in the range $[-\delta r_0\Omega_0,\delta r_0\Omega_0]$ for each
velocity component, while the magnetic field perturbations vanish
$\tilde{\vek b}_{\vek
  1}=0$. 

In simulations \#SCA15a--\#SCA16c and \#SCR15a--c from
\tabref{tab:Snoopy_channels}, the initial velocity and magnetic field
are given by
\begin{align}
  {\vek v} &= -q \Omega_0 (r - r_0) {\bm{ \hat{\phi}}}+  \bfvc  + \tilde{\vek v}_{\vek 1} ,  
\label{eq:channel_full_Snoopy_v}
\\
  {\vek b} &=  b_{0z} {\vek{\hat{z}}} +     \bfbc      +    {\vek b_1} ,
\label{eq:channel_full_Snoopy_b}
\end{align}
where $ \bfvc$ and $ \bfbc$ are the full channel mode solution given
by \eqsref{eq:channel_v} and (\ref{eq:channel_b}), respectively, and
$\tildevc = \epsilon \Omega_0 r_0 $ 
and $\tildebc$ is determined from \Eqref{eq:amplitude_ratio}.  In
simulations \#SCR15a--c random perturbations $\tilde{\vek v}_{\vek 1}$
and $\tilde{\vek b}_{\vek 1}$ follow equation~(\ref{eq:Snoopy_random})
as explained above, whereas in simulations \#SCA15a--\#SCA16c they are
replaced by cell-by-cell random values for the velocity and zero
magnetic field perturbation.

\vlc{\citet{Rembiasz_et_al} comprehensively studied the influence
  of the geometry of the computational domain, i.e.\, its aspect
  ratio both in the vertical and azimuthal direction, on the magnetic
  field amplification.  Based on these studies, we have chosen
  computational boxes in such a way as to not affect the MRI
  termination amplitude.  }

\section{Results}
\label{sec:results}

\begin{table*}
  \caption[]{Overview of our 3D MRI simulations done with \textsc{Aenus}.
    The columns give the model identifier, the magnetic field
    strength, the hydrodynamic and magnetic Reynolds numbers ($\Ree$ and $\Rm$,
    respectively),  MRI wavelength, $\lambdamri$, 
    the size of the computational domain, the resolution, the 
    number of grid cells per MRI wavelength, initial perturbation
    amplitudes $\epsilon$ and $\delta$ (see Eq.~\ref{eq:v_with_sin}) ,
    the volume-averaged Maxwell stress (Eq.~\ref{eq:MMM}) at termination,
    and the amplification factor (Eq.~\ref{eq:AAA}).
  }
%
\begin{center}
\begin{tabular}{|c|l|c|c|c|c|c|c|c|l|c|c|c|c|c|c|}
\hline
 \#& \pbox{5cm}{$b_{0z}$ \\ $[10^{13}$ G$]$ } & $\Ree$ & $\Rm$ &
   $\lambdamri [\km]$ &       \pbox{5cm}{box size \\ $(r \times \phi \times z)$ [km]} & 
      \pbox{5cm}{resolution \\ $(r \times \phi \times z)$ \ } & 
      \pbox{5cm}{zones per \\ channel } & $\epsilon [10^{-5}]$ &  $\delta [10^{-5}]$ & 
      \pbox{5cm}{$\MMMterm$ \\ $[10^{28}\ \mathrm{G}^2]$}  & $\AAA$
\\  \hline
   A1 & \hspace{0.3cm} 0.73 & $\infty$ & 100 & 0.4 &$2 \times 2 \times 0.4$ & $100 \times 100 \times 20$ & $ 20$ & $0.2$ & $1$& 3.3  & 24.8
 \\
   A2 & \hspace{0.3cm} 0.73 &$\infty$ & 100 & 0.4 & $2 \times 2 \times 0.4$      & $168 \times 168 \times 34$ & $ 34$ & $0.2$ & $1$& 2.6 &  22.2
 \\
   A3 & \hspace{0.3cm} 0.73 & $\infty$ & 100 & 0.4 & $2 \times 2 \times 0.4$      & $ 336 \times 336 \times 68$ & $ 68$ & $0.2$ & $1$& 2.5  & 21.4
 \\
   A4 & \hspace{0.3cm} 0.73 & $\infty$ & 100 & 0.4 & $2 \times 2 \times 0.4$      & $672 \times 672 \times 136$ & $ 136$ & $0.2$ & $1$& 2.5 & 21.3
 \\
\hline
   A5 & \hspace{0.3cm} 0.92 & $\infty$ &  100 &0.5 & $2 \times 2 \times 0.5$     & $136 \times 136 \times 34$ & $ 34$ & $0.2$ & $1$& 3.5  & 20.3
 \\
   A6 & \hspace{0.3cm} 0.92 & $\infty$ &  100 & 0.5 & $2 \times 2 \times 0.5$      & $272 \times 272 \times 68$ & $ 68$ & $0.2$ & $1$& 3.4  & 20.0
 \\
\hline
   A7 & \hspace{0.3cm} 1.22 & $\infty$ & 100 & 0.666 &$2 \times 8 \times 2$          & $ 60 \times 240 \times  60$ & $ 20$ & $0.2$ & $1$ & 7.8  & 22.9 
\\
   A8a & \hspace{0.3cm} 1.22 & $\infty$ & 100 & 0.666 & $2 \times 8 \times 2$          & $100 \times 400 \times 100$ & $ 33$ & $0.2$ & $1$ & 6.6  & 21.0
 \\
   A8b & \hspace{0.3cm} 1.22 & $\infty$ & 100 & 0.666 & $2 \times 8 \times 2$          & $100 \times 400 \times 100$ & $ 33$ & $0.2$ & $1$ & 6.6 & 21.0
 \\
   A8c & \hspace{0.3cm} 1.22 & $\infty$ & 100 & 0.666 & $2 \times 8 \times 2$          & $100 \times 400 \times 100$ & $ 33$ & $0.2$ & $1$ & 6.6 & 21.1
 \\
   A9a & \hspace{0.3cm} 1.22 & $\infty$ & 100 & 0.666 & $2 \times 2 \times 0.666$      & $100 \times 100 \times  34$ & $ 34$ & $0.2$ & $1$ & 6.6 & 21.0
 \\
  A9b & \hspace{0.3cm} 1.22 & $\infty$ & 100 & 0.666 & $2 \times 2 \times 0.666$     & $100 \times 100 \times  34$ & $ 34$ & $0.2$ & $1$ & 6.0 & 20.0
 \\
  A10 & \hspace{0.3cm} 1.22 & $\infty$ & 100 & 0.666 & $2 \times 8 \times 2$          & $200 \times 800 \times 200$ & $ 67$ & $0.2$ & $1$ & 5.2 & 18.6
 \\
  A11 & \hspace{0.3cm} 1.22 & $\infty$ & 100 & 0.666 & $2 \times 2 \times 0.666$      & $400 \times 400 \times 134$ & $134$ & $0.2$ & $1$& 5.2 & 18.6
 \\
\hline
  \end{tabular}
\label{tab:aenus}
\end{center}
\begin{flushleft}
\footnotesize{$^{\ast}\,$ \cross{Note  that} $\lambdamri$, $\Ree$ and $\Rm$ are not
    constant in the computational domain, but vary by $\approx 20\%$
\citep[see][for a discussion]{Rembiasz_et_al}.}\\
\footnotesize{\vlc{$^{\ast\ast }\,$$\Ree=\infty$
    means that the only viscosity present in the models is of
    numerical origin. The numerical viscosity of the code is discussed
    in \cite{Rembiasz}. } }
\end{flushleft}
 \end{table*}

\begin{table*}
  \caption[]{Overview of our 3D MRI simulations done with
    \textsc{Snoopy}.
The columns are like in \tabref{tab:aenus}  but the initial perturbations are defined in equations (\ref{eq:Snoopy_random})--(\ref{eq:channel_full_Snoopy_b}).
}
%
\begin{center}
\begin{tabular}{|c|l|c|c|c|c|c|c|c|l|c|c|c|c|c|c|}
\hline
 \#& \pbox{5cm}{$b_{0z}$ \\ $[10^{13}$ G$]$ } & $\Ree$ & $\Rm$ &
   $\lambdamri [\km]$ &       \pbox{5cm}{box size \\ $(r \times \phi \times z)$ [km]} & 
      \pbox{5cm}{resolution \\ $(r \times \phi \times z)$ \ } & 
      \pbox{5cm}{zones per \\ channel } & $\epsilon [10^{-5}]$ &$\delta [10^{-5}]$ &
      \pbox{5cm}{$\MMMterm$ \\ $[10^{28}\ \mathrm{G}^2]$}  &$\AAA$ 
\\  \hline
   S1a & \hspace{0.3cm} 0.12 & 100 & 100 & 0.067 &$2 \times 2 \times 0.067$ & $1920 \times 960 \times 64$ & $ 64 $  &$0.02$ & $0.1$ & $0.39$  & 53.2
 \\
   S1b & \hspace{0.3cm} 0.12 & 100 & 100 & 0.067 &$2 \times 2 \times 0.067$ & $1920 \times 960 \times 64$ & $ 64 $  & $0.02$ & $0.1$ & $0.34$  & 49.3
 \\
   S1c & \hspace{0.3cm} 0.12 & 100 & 100 & 0.067 &$2 \times 2 \times 0.067$ & $1920 \times 960 \times 64$ & $ 64 $  & $0.02$ & $0.1$ & $0.27$  & 44.0
 \\
   S1d & \hspace{0.3cm} 0.12 & 100 & 100 & 0.067 &$2 \times 2 \times 0.067$ & $1920 \times 960 \times 64$ & $ 64 $  & $0.02$ & $0.1$ & $0.32$  & 48.2
 \\
   S1e & \hspace{0.3cm} 0.12 & 100 & 100 & 0.067 &$2 \times 2 \times 0.067$ & $1920 \times 960 \times 64$ & $ 64 $  & $0.02$ & $0.1$ & $0.35$  & 50.0
 \\
\hline
  S2a & \hspace{0.3cm} 0.24 & 100 & 100 & 0.133 &$2 \times 2 \times 0.133$ & $960 \times 480 \times 64$ & $ 64 $ & $0.04$ & $0.2$ & $1.31$  & 48.8
 \\ 
  S2b & \hspace{0.3cm} 0.24 & 100 & 100 & 0.133 &$2 \times 2 \times 0.133$ & $960 \times 480 \times 64$ & $ 64 $  & $0.04$ & $0.2$ & $1.47$  & 51.6
 \\ 
  S2c & \hspace{0.3cm} 0.24 & 100 & 100 & 0.133 &$2 \times 2 \times 0.133$ & $960 \times 480 \times 64$ & $ 64 $  & $0.04$ & $0.2$ & $1.55$  & 52.9
 \\ 
  S2d & \hspace{0.3cm} 0.24 & 100 & 100 & 0.133 &$2 \times 2 \times 0.133$ & $960 \times 480 \times 64$ & $ 64 $  & $0.04$ & $0.2$ & $1.78$  & 56.8
 \\ 
  S2e & \hspace{0.3cm} 0.24 & 100 & 100 & 0.133 &$2 \times 2 \times 0.133$ & $960 \times 480 \times 64$ & $ 64 $  & $0.04$ & $0.2$ & $1.24$  & 47.5
 \\ 
\hline
  S3 & \hspace{0.3cm} 0.59 & 100 & 100 & 0.333 &$2 \times 2 \times 0.333$ & $384 \times 192 \times 64$ & $ 64 $  & $0.1$ & $0.5$ & $11.77$  & 58.4
\\
  S4 & \hspace{0.3cm} 0.59 & 100 & 100 & 0.333 &$2 \times 2 \times 0.333$ & $768 \times 384 \times 128$ & $ 128 $ & $0.1$ & $0.5$ & $5.88$  & 41.3
\\
\hline
  S5 & \hspace{0.3cm} 0.64 & 100 & 100 & 0.364 &$2 \times 2 \times 0.364$ & $352 \times 176 \times 64$ & $ 64 $  & $0.11$ & $0.55$ & $11.37$  & 52.6
\\
  S6 & \hspace{0.3cm} 0.64 & 100 & 100 & 0.364 &$2 \times 2 \times 0.364$ & $704 \times 352 \times 128$ & $ 128 $ & $0.11$ & $0.55$ & $9.07$  & 47.0
\\
\hline
  S7a & \hspace{0.3cm} 0.71 & 100 & 100 & 0.400&$2 \times 2 \times 0.400$ & $320 \times 160 \times 64$ & $ 64 $  & $0.12$ & $0.60$ & $12.33$  & 49.8
\\
  S7b & \hspace{0.3cm} 0.71 & 100 & 100 & 0.400 &$2 \times 2 \times 0.400$ & $320 \times 160 \times 64$ & $ 64 $  & $0.12$ & $0.60$ & $20.27$  & 63.9
\\
  S7c & \hspace{0.3cm} 0.71 & 100 & 100 & 0.400 &$2 \times 2 \times 0.400$ & $320 \times 160 \times 64$ & $ 64 $  & $0.12$ & $0.60$ & $16.87$  & 58.2
\\
  S7d & \hspace{0.3cm} 0.71 & 100 & 100 & 0.400 &$2 \times 2 \times 0.400$ & $320 \times 160 \times 64$ & $ 64 $  & $0.12$ & $0.60$ & $16.02$  & 56.8
\\
  S7e & \hspace{0.3cm} 0.71 & 100 & 100 & 0.400 &$2 \times 2 \times 0.400$ & $320 \times 160 \times 64$ & $ 64 $  & $0.12$ & $0.60$ & $9.66$  & 44.1
\\
  S8a & \hspace{0.3cm} 0.71 & 100 & 100 & 0.400 &$2 \times 2 \times 0.400$ & $640 \times 320 \times 128$ & $ 128 $  &$0.12$ & $0.60$ & $19.13$  & 62.0
\\
  S8b & \hspace{0.3cm} 0.71 & 100 & 100 & 0.400 &$2 \times 2 \times 0.400$ & $640 \times 320 \times 128$ & $ 128 $  & $0.12$ & $0.60$ & $13.40$  & 51.9
\\
  S8c & \hspace{0.3cm} 0.71 & 100 & 100 & 0.400 &$2 \times 2 \times 0.400$ & $640 \times 320 \times 128$ & $ 128 $  & $0.12$ & $0.60$ & $11.92$  & 49.0
\\
  S8d & \hspace{0.3cm} 0.71 & 100 & 100 & 0.400 &$2 \times 2 \times 0.400$ & $640 \times 320 \times 128$ & $ 128 $  & $0.12$ & $0.60$ & $21.76$  & 66.2
\\
  S8e & \hspace{0.3cm} 0.71 & 100 & 100 & 0.400 &$2 \times 2 \times 0.400$ & $640 \times 320 \times 128$ & $ 128 $  & $0.12$ & $0.60$ & $13.89$  & 52.9
\\
\hline
  S9 & \hspace{0.3cm} 0.78 & 100 & 100 &  0.444 &$2 \times 2 \times 0.444$ & $288 \times 144 \times 64$ & $ 64 $  & $0.13$ & $0.67$ & $20.4$  & 57.7
\\
  S10 & \hspace{0.3cm} 0.78 & 100 & 100 &  0.444 &$2 \times 2 \times 0.444$ & $576 \times 288 \times 128$ & $ 128 $  & $0.13$ & $0.67$ & $13.02$  & 46.0
\\
\hline 
  S11 & \hspace{0.3cm} 0.88 & 100 & 100 &  0.500&$2 \times 2 \times 0.500$ & $256 \times 128 \times 64$ & $ 64 $  & $0.15$ & $0.75$ & $20.75$  & 51.7
\\
  S12 & \hspace{0.3cm} 0.88 & 100 & 100 &  0.500 &$2 \times 2 \times 0.500$ & $512 \times 256 \times 128$ & $ 128 $  & $0.15$ & $0.75$ & $18.38$  & 48.6
\\
\hline 
  S13a & \hspace{0.3cm} 1.01 & 100 & 100 &  0.571 &$2 \times 2 \times 0.571$ & $224 \times 112 \times 64$ & $ 64 $  & $0.17$ & $0.86$ & $34.74$  & 58.5
\\
  S13b & \hspace{0.3cm} 1.01 & 100 & 100 &  0.571 &$2 \times 2 \times 0.571$ & $224 \times 112 \times 64$ & $ 64 $  & $0.17$ & $0.86$ & $56.91$  & 74.9
\\
  S13c & \hspace{0.3cm} 1.01 & 100 & 100 &  0.571 &$2 \times 2 \times 0.571$ & $224 \times 112 \times 64$ & $ 64 $  & $0.17$ & $0.86$ & $32.15$  & 56.3
\\
  S13d & \hspace{0.3cm} 1.01 & 100 & 100 &  0.571 &$2 \times 2 \times 0.571$ & $224 \times 112 \times 64$ & $ 64 $  & $0.17$ & $0.86$ & $25.85$  & 50.5
\\
  S13e & \hspace{0.3cm} 1.01 & 100 & 100 &  0.571 &$2 \times 2 \times 0.571$ & $224 \times 112 \times 64$ & $ 64 $  & $0.17$ & $0.86$ & $64.49$  & 79.7
\\
  S14a & \hspace{0.3cm} 1.01 & 100 & 100 &  0.571 &$2 \times 2 \times 0.571$ & $448 \times 224 \times 128$ & $ 128 $  &$0.17$ & $0.86$ & $50.40$  & 70.5
\\
  S14b & \hspace{0.3cm} 1.01 & 100 & 100 &  0.571 &$2 \times 2 \times 0.571$ & $448 \times 224 \times 128$ & $ 128 $  & $0.17$ & $0.86$ & $36.77$  & 60.2
\\
  S14c & \hspace{0.3cm} 1.01 & 100 & 100 &  0.571 &$2 \times 2 \times 0.571$ & $448 \times 224 \times 128$ & $ 128 $  & $0.17$ & $0.86$ & $34.17$  & 58.0
\\
  S14d & \hspace{0.3cm} 1.01 & 100 & 100 &  0.571 &$2 \times 2 \times 0.571$ & $448 \times 224 \times 128$ & $ 128 $  & $0.17$ & $0.86$ & $37.22$  & 60.6
\\
  S14e & \hspace{0.3cm} 1.01 & 100 & 100 &  0.571 &$2 \times 2 \times 0.571$ & $448 \times 224 \times 128$ & $ 128 $  & $0.17$ & $0.86$ & $38.95$ & 62.0
\\
\hline
  S15a & \hspace{0.3cm} 1.18 & 100 & 100 &  0.666&$2 \times 2 \times 0.666$ & $192 \times 96 \times 64$ & $ 64 $  & $0.2$ & $0.1$ & $164.56$  & 109.2
\\
  S15b & \hspace{0.3cm} 1.18 & 100 & 100 &  0.666 &$2 \times 2 \times 0.666$ & $192 \times 96 \times 64$ & $ 64 $  & $0.2$ & $0.33$ & $105.31$  & 87.3
\\
  S15c & \hspace{0.3cm} 1.18 & 100 & 100 &  0.666 &$2 \times 2 \times 0.666$ & $192 \times 96 \times 64$ & $ 64 $  & $0.2$ & $1$ & $46.53$  & 58.0
\\
  S15d & \hspace{0.3cm} 1.18 & 100 & 100 &  0.666 &$2 \times 2 \times 0.666$ & $192 \times 96 \times 64$ & $ 64 $  & $0.2$ & $1$ & $53.33$  & 62.1
\\
  S15e & \hspace{0.3cm} 1.18 & 100 & 100 &  0.666 &$2 \times 2 \times 0.666$ & $192 \times 96 \times 64$ & $ 64 $  & $0.67$ & $1$ & $73.53$  & 73.0
\\
  S15f & \hspace{0.3cm} 1.18 & 100 & 100 &  0.666 &$2 \times 2 \times 0.666$ & $192 \times 96 \times 64$ & $ 64 $  & $2$ & $1$ & $77.67$  & 75.0
\\
  S15g & \hspace{0.3cm} 1.18 & 100 & 100 &  0.666 &$2 \times 2 \times 0.666$ & $192 \times 96 \times 64$ & $ 64 $  & $6.67$ & $1$ & $95.09$  & 83.0
\\
  S15h & \hspace{0.3cm} 1.18 & 100 & 100 &  0.666 &$2 \times 2 \times 0.666$ & $192 \times 96 \times 64$ & $ 64 $  & $20$ & $1$ & $91.70$  & 81.5
\\
  S16a & \hspace{0.3cm} 1.18 & 100 & 100 &  0.666 &$2 \times 2 \times 0.666$ & $384 \times 192 \times 128$ & $ 128 $  & $0.2$ & $0.1$ & $146.86$  & 103.1
\\
  S16b & \hspace{0.3cm} 1.18 & 100 & 100 &  0.666 &$2 \times 2 \times 0.666$ & $384 \times 192 \times 128$ & $ 128 $  & $0.2$ & $0.33$ & $95.08$  & 83.0
\\
  S16c & \hspace{0.3cm} 1.18 & 100 & 100 &  0.666 &$2 \times 2 \times 0.666$ & $384 \times 192 \times 128$ & $ 128 $  & $0.2$ & $1$ & $47.83$  & 58.8
\\
  S16d & \hspace{0.3cm} 1.18 & 100 & 100 &  0.666 &$2 \times 2 \times 0.666$ & $384 \times 192 \times 128$ & $ 128 $  & $0.2$ & $1$ & $45.17$ & 57.2
\\
  S16e & \hspace{0.3cm} 1.18 & 100 & 100 &  0.666 &$2 \times 2 \times 0.666$ & $384 \times 192 \times 128$ & $ 128 $  & $0.67$ & $1$ & $58.62$ & 65.1
\\
  S16f & \hspace{0.3cm} 1.18 & 100 & 100 &  0.666 &$2 \times 2 \times 0.666$ & $384 \times 192 \times 128$ & $ 128 $  & $2$ & $1$ & $69.53$  & 70.9
\\
  S16g & \hspace{0.3cm} 1.18 & 100 & 100 &  0.666 &$2 \times 2 \times 0.666$ & $384 \times 192 \times 128$ & $ 128 $  & $6.67$ & $1$ & $82.19$  & 82.2
\\
  S16h & \hspace{0.3cm} 1.18 & 100 & 100 &  0.666 &$2 \times 2 \times 0.666$ & $384 \times 192 \times 128$ & $ 128 $ &$20$ & $1$ & $97.67$  & 84.1
\\
  S17a & \hspace{0.3cm} 1.18 & 100 & 100 &  0.666 &$2 \times 2 \times 0.666$ & $768 \times 384 \times 256$ & $ 256 $ & $0.2$ & $1$ & $56.11$ & 63.7
\\
  S17b & \hspace{0.3cm} 1.18 & 100 & 100 &  0.666 &$2 \times 2 \times 0.666$ & $768 \times 384 \times 256$ & $ 256 $ & $0.2$ & $1$ & $54.63$ & 62.9
\\
  S18 & \hspace{0.3cm} 1.18 & 100 & 100 &  0.666 &$2 \times 2 \times 0.666$ & $1536 \times 768 \times 512$ & $ 512 $ & $0.2$ & $1$ & $32.70$  & 48.7
\\
  S19 & \hspace{0.3cm} 1.18 & 100 & 100 &  0.666 &$2 \times 2 \times 0.666$ & $192 \times 96 \times 64$ & $ 64$ & $0.2$ & $1$ & $101.58$  & 84.2
\\
\hline
  \end{tabular}
\label{tab:Snoopy}
\end{center}
 \end{table*}

\begin{table*}
  \contcaption{Overview of our 3D MRI simulations done with
    \textsc{Snoopy}, perturbations like in \textsc{Aenus}
    The columns are like in \tabref{tab:aenus} but the initial 
    perturbations are defined in equations 
    (\ref{eq:Snoopy_random})--(\ref{eq:channel_full_Snoopy_b}).
  }
%
\begin{center}
\begin{tabular}{|c|l|c|c|c|c|c|c|c|l|c|c|c|c|c|c|c|}
\hline
 \#& \pbox{5cm}{$b_{0z}$ \\ $[10^{13}$ G$]$ } & $\Ree$ & $\Rm$ &
   $\lambdamri [\km]$ &       \pbox{5cm}{box size \\ $(r \times \phi \times z)$ [km]} & 
      \pbox{5cm}{resolution \\ $(r \times \phi \times z)$ \ } & 
      \pbox{5cm}{zones per \\ channel } & $\epsilon [10^{-5}]$ &$\delta [10^{-5}]$ &
      \pbox{5cm}{ $\MMMterm$ \\ $[10^{28}\ \mathrm{G}^2]$}  &$\AAA$ 
\\  \hline
  SA15a & \hspace{0.3cm} 1.18 & 100 & 100 &  0.666&$2 \times 2 \times 0.666$ & $192 \times 96 \times 64$ & $ 64 $ &$0.2$ & $1$ &$116.83$  & 91.6
\\
  SA15b & \hspace{0.3cm} 1.18 & 100 & 100 &  0.666&$2 \times 2 \times 0.666$ & $192 \times 96 \times 64$ & $ 64 $ & $0.2$ & $1$ & $106.44$  & 87.4
\\
  SA15c & \hspace{0.3cm} 1.18 & 100 & 100 &  0.666&$2 \times 2 \times 0.666$ & $192 \times 96 \times 64$ & $ 64 $ & $0.2$ & $1$ & $118.24$  & 88.6
\\
  SA15d & \hspace{0.3cm} 1.18 & 100 & 100 &  0.666&$2 \times 2 \times 0.666$ & $192 \times 96 \times 64$ & $ 64 $ & $0.2$ & $1$ & $118.24$  & 92.2
\\
  SA15e & \hspace{0.3cm} 1.18 & 100 & 100 &  0.666&$2 \times 2 \times 0.666$ & $192 \times 96 \times 64$ & $ 64 $ & $0.2$ & $1$ & $105.71$  & 87.1
\\  \hline
  SCA15a & \hspace{0.3cm} 1.18 & 100 & 100 &  0.666&$2 \times 2 \times 0.666$ & $192 \times 96 \times 64$ & $ 64 $ & $1233$ &$22.7$ & $35.52$  & 50.5
\\
  SCA15b & \hspace{0.3cm} 1.18 & 100 & 100 &  0.666&$2 \times 2 \times 0.666$ & $192 \times 96 \times 64$ & $ 64 $ & $1233$ & $227$ & $25.85$  & 43.1
\\
  SCA15c & \hspace{0.3cm} 1.18 & 100 & 100 &  0.666&$2 \times 2 \times 0.666$ & $192 \times 96 \times 64$ & $ 64 $ & $1233$ & $227$ & $24.68$  & 42.1
\\
  SCA15d & \hspace{0.3cm} 1.18 & 100 & 100 &  0.666&$2 \times 2 \times 0.666$ & $192 \times 96 \times 64$ & $ 64 $ & $1233$ & $227$ & $23.28$  & 40.9
\\
  SCA15e & \hspace{0.3cm} 1.18 & 100 & 100 &  0.666&$2 \times 2 \times 0.666$ & $192 \times 96 \times 64$ & $ 64 $ & $1233$ & $2270$ & $10.00$  & 26.8
\\
  SCA15f & \hspace{0.3cm} 1.18 & 100 & 100 &  0.666&$2 \times 2 \times 0.666$ & $192 \times 96 \times 64$ & $ 64 $ & $1233$ & $2270$ & $10.21$  & 27.1
\\
  SCA15g & \hspace{0.3cm} 1.18 & 100 & 100 &  0.666&$2 \times 2 \times 0.666$ & $192 \times 96 \times 64$ & $ 64 $ & $1233$ & $2270$ & $8.64$  & 24.9
\\
\hline 
  SCR15a & \hspace{0.3cm} 1.18 & 100 & 100 &  0.666&$2 \times 2 \times 0.666$ & $192 \times 96 \times 64$ & $ 64 $ & $1233$ & $227$ & $5.78$  & 20.4
\\
  SCR15b & \hspace{0.3cm} 1.18 & 100 & 100 &  0.666&$2 \times 2 \times 0.666$ & $192 \times 96 \times 64$ & $ 64 $ & $1233$ & $227$ & $4.44$  & 17.9
\\
  SCR15c & \hspace{0.3cm} 1.18 & 100 & 100 &  0.666&$2 \times 2 \times 0.666$ & $192 \times 96 \times 64$ & $ 64 $ & $1233$ & $227$ & $7.25$  & 22.8
\\
SCA16a & \hspace{0.3cm} 1.18 & 100 & 100 &  0.666 &$2 \times 2 \times 0.666$ & $384 \times 192 \times 128$ & $ 128 $ & $1233$ & $227$ & $31.80$  & 47.8
\\
SCA16b & \hspace{0.3cm} 1.18 & 100 & 100 &  0.666 &$2 \times 2 \times 0.666$ & $384 \times 192 \times 128$ & $ 128 $ & $1233$ & $227$ & $30.17$  & 46.6
\\
SCA16c & \hspace{0.3cm} 1.18 & 100 & 100 &  0.666 &$2 \times 2 \times 0.666$ & $384 \times 192 \times 128$ & $ 128 $ & $1233$ & $227$ & $35.03$  & 50.2
\\
\hline
  \end{tabular}
\label{tab:Snoopy_channels}
\end{center}
 \end{table*}


\subsection{Termination criterion}

The goal of this section is to determine which termination criterion
(I or II, Eqs.~\ref{eq:term1} and \ref{eq:term2}, respectively) is a
better predictor for the MRI termination.  To this end, we analyse in
detail simulations \#A8a \citep[which was also presented in][as model
\#7, see Appendix \ref{app:conversion}, for details]{Rembiasz_et_al}
and \#S19 performed with \textsc{Aenus} and \textsc{Snoopy},
respectively.

\subsubsection{Aenus simulation}
\label{sususeAe}

We have calculated spatial discrete 3D Fourier transforms of the
magnetic field, velocity and ${\vek w } \equiv \sqrt{\rho} {\vek v}$
components $b_\beta$, $v_\beta$, and $w_\beta$, respectively, with
$\beta \in \{ r, \phi, z \} $ at a given time using a Fast Fourier
Transform (FFT) algorithm.  We denote the complex FFT coefficients
with $\hat{b}_\beta$, $\hat{v}_\beta$ and $\hat{w}_\beta$,
respectively.  The power spectral density of the magnetic field is
proportional to $|\hat{b}_\beta|^2$, which is a measure of the average
magnetic field energy density of the component $b_\beta$ in Fourier
space.  Analogously, the kinetic energy density is proportional to
$|\hat{w}_\beta|^2$.  We expect that MRI channel flows appear as
structures with a wavevector
\begin{equation}
  {\bf \kmri} = (0, 0, \kmri),
\end{equation}
and the parasitic instabilities develop with non-zero radial and
azimuthal wavenumbers.

The average magnetic and kinetic energy density of the field
components $b_\beta$ and $w_\beta$ can be computed from the Fourier
amplitudes as
\begin{align}
  e^{{\rm mag}}_{\beta} &=  \frac{1}{2}
                           \sum_{l = -N_r/2}^{ N_r/2} \,  
                                     \sum_{m = -N_\phi/2}^{N_\phi/2}\,  
                                     \sum_{n = -N_z/2}^{N_z/2}
                         |\hat{b}_\beta(k_{l}, k_{m},k_{n})|^2  \text{\ \  and} \\
  e^{{\rm kin}}_{ \beta} &=  
                             \frac{1}{2}
                            \sum_{l = -N_r/2}^{ N_r/2} \,  
                                     \sum_{m = -N_\phi/2}^{N_\phi/2}\,  
                                     \sum_{n = -N_z/2}^{N_z/2}
                         |\hat{w}_\beta(k_{l}, k_{m},k_{n})|^2,
\end{align}
where
\begin{equation}
  (k_{l}, k_{m},k_{n}) = \left( \frac{2\pi \,l}{L_r}, 
                               \frac{2 \pi \,m}{L_\phi},
                               \frac{2 \pi \,n}{L_z} \right).
\end{equation}
We can estimate the average magnetic energy density of the
MRI channels
restricting the summation to locations in Fourier space relevant for 
this
instability, i.e.
\begin{align}
\label{eq:mri_mag}
  e^{{\rm mag}}_{{\rm MRI}, \alpha} &=  |\hat{b}_\alpha(0, 0, k_{\rm MRI})|^2,
\\
\label{eq:mri_kin}
  e^{{\rm kin}}_{{\rm MRI}, \alpha} &=  |\hat{w}_\alpha(0, 0, k_{\rm MRI})|^2,
\end{align}
where here and in the following the subscript $\alpha$ is restricted
to $\alpha \in \{r,\phi\}$.

To determine the horizontal component of the wavevector of the
parasitic instabilities, \citet{Rembiasz_et_al} analysed Fourier modes
of $b_\alpha$ with finite $k_{r}$ and $k_{\phi}$, but $k_z = 0$.  They
found that the parasitic instabilities produce a characteristic
signature with wavevectors $ {\bf k}_{\rm p} = (k_r, k_\phi, 0) $,
where $k_r \simeq k_\phi \simeq 0.8 \kmri$.  This means that in
accordance with the parasitic model, parasitic instabilities develop
along the velocity channels.

In this paper, we want to find an estimator for the total magnetic and
kinetic energy stored in the parasitic instabilities.  It is clear
however, that the parasites will also contribute to other Fourier
modes. For instance, according to \citet{Pessah}, the dominant
parasitic mode will contribute to all Fourier modes with
$ {\bf k}_{\rm p} = (\zeta \kmri, \zeta \kmri, n \kmri)$, where
$\zeta = 0.42$ and $n$ is a natural number.  However, we expect to see
not only the dominant modes in our simulations. Moreover, as pointed
out by \citet{Latter_et_al_2}, PG09 and \citet{Pessah} the rotational
shear will modify the horizontal components of the parasitic modes
(making them time dependent). Thus, we take
\begin{align}
\label{eq:p_mag}
  e^{{\rm mag}}_{{\rm p}, \beta}  &= e^{{\rm mag}}_{ \beta}  -
                         \frac{1}{2}
                                     \sum_{l = -N_r/2}^{N_r/2}\,  
                          \sum_{n= -N_z/2}^{N_z/2}  
                          |\hat{b}_\beta(k_{l}, 0, k_{n})|^2 \\
\label{eq:p_kin}
  e^{{\rm kin}}_{{\rm p}, \beta}   &= e^{{\rm kin}}_{\beta}   -
                           \frac{1}{2} 
                                       \sum_{l = -N_r/2}^{N_r/2}\,  
                          \sum_{n= -N_z/2}^{N_z/2}   
                          |\hat{w}_\beta(k_{l}, 0, k_{n})|^2,
\end{align}
as an estimator for the energy stored in parasitic modes, i.e.\ we
take all Fourier modes but the axisymmetric ones.  

\cross{We exclude the axisymmetric modes because they not only
  include the MRI channels, but also contributions due to: (i)
  $\hat{w}_{\phi}-$modes arising from the background differential
  rotation; (ii) radial ($\hat{w}_{r}-$) modes possibly excited by
  (small) deviations from hydrodynamical equilibrium caused by the
  numerical inaccuracies; (iii) modes driven by the radial boundary
  conditions.}

\begin{figure*}
\centering
\includegraphics[width=0.48\linewidth]{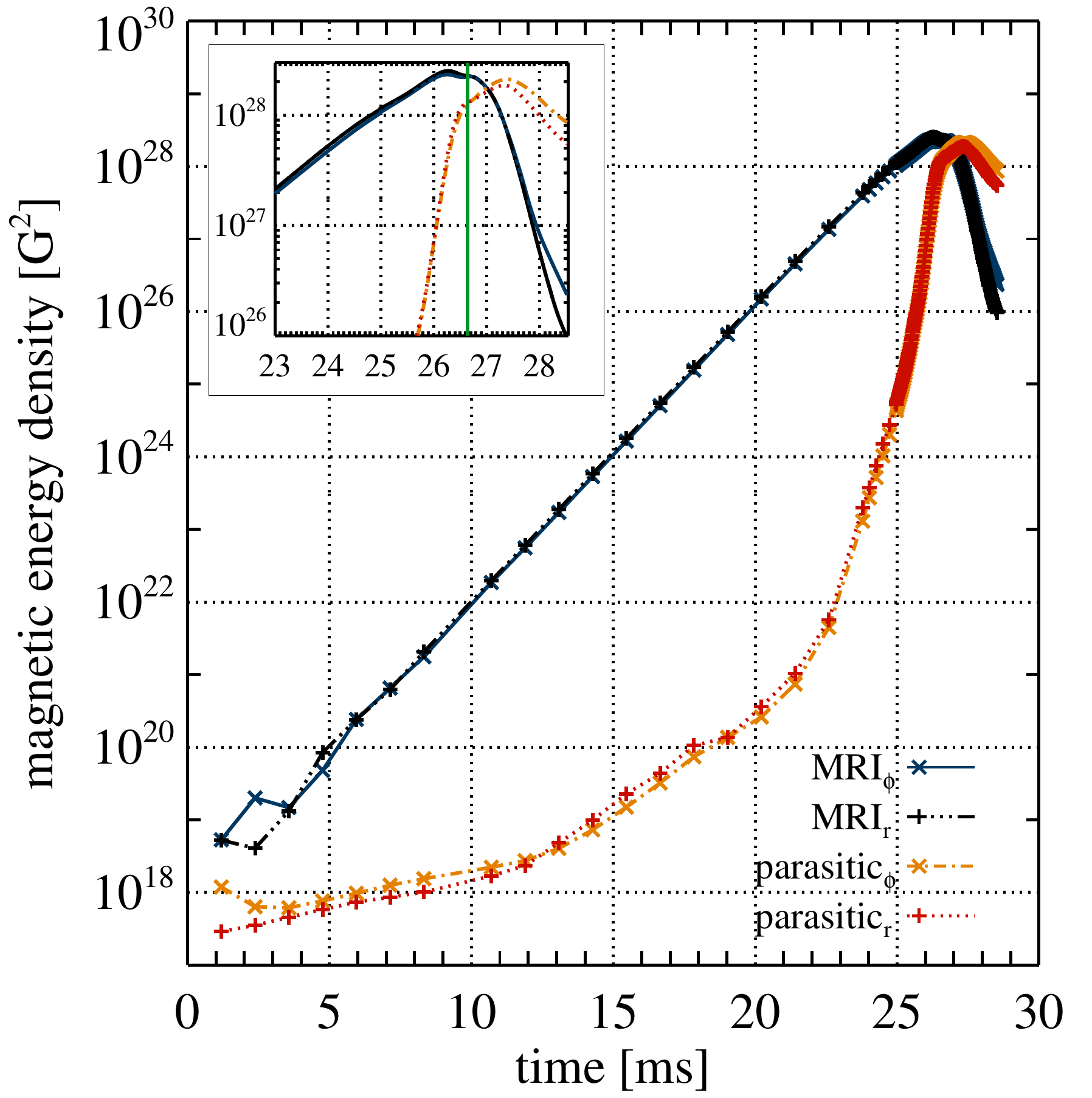}
\includegraphics[width=0.48\linewidth]{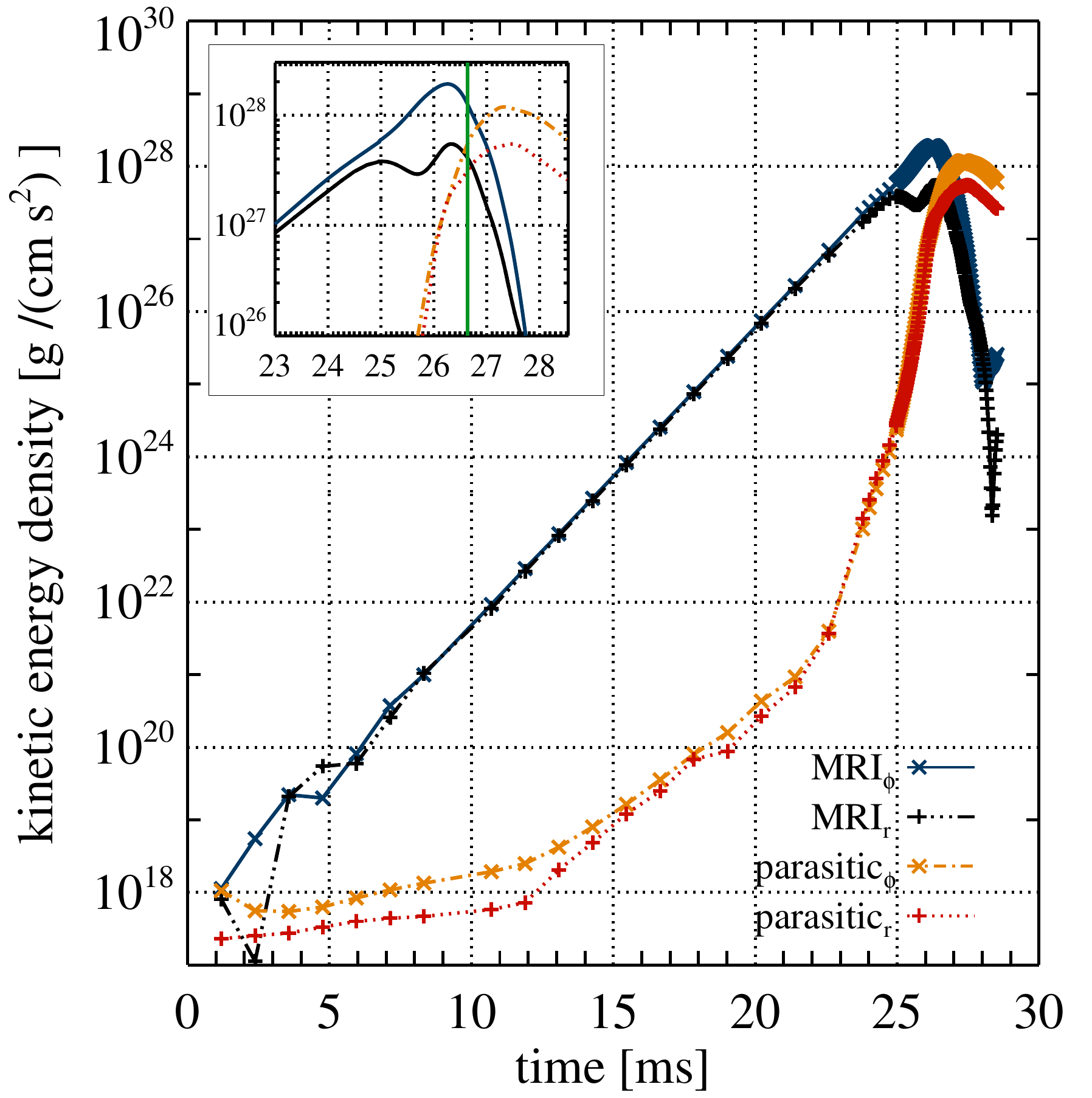}
\includegraphics[width=0.48\linewidth]{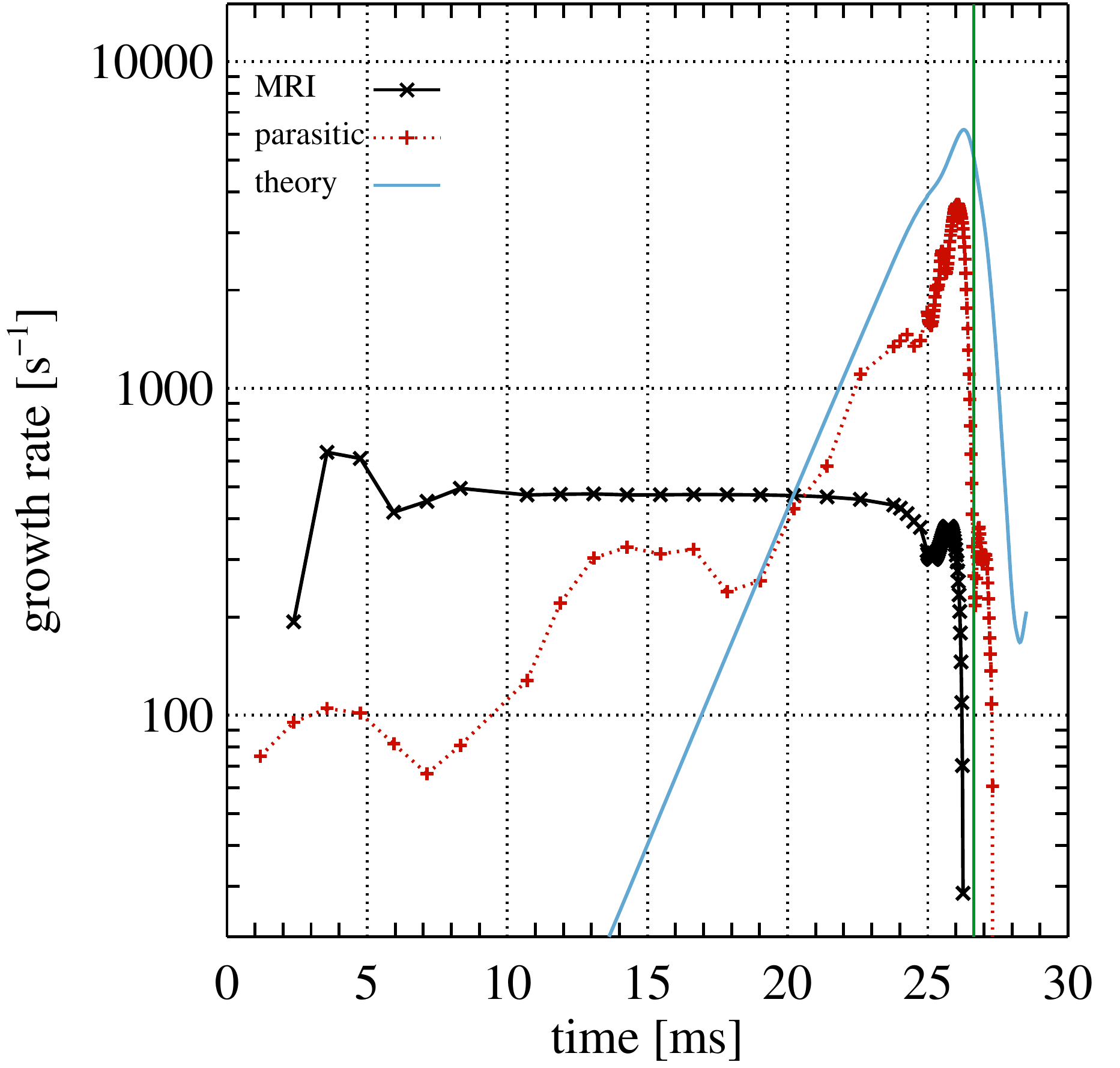}
\includegraphics[width=0.48\linewidth]{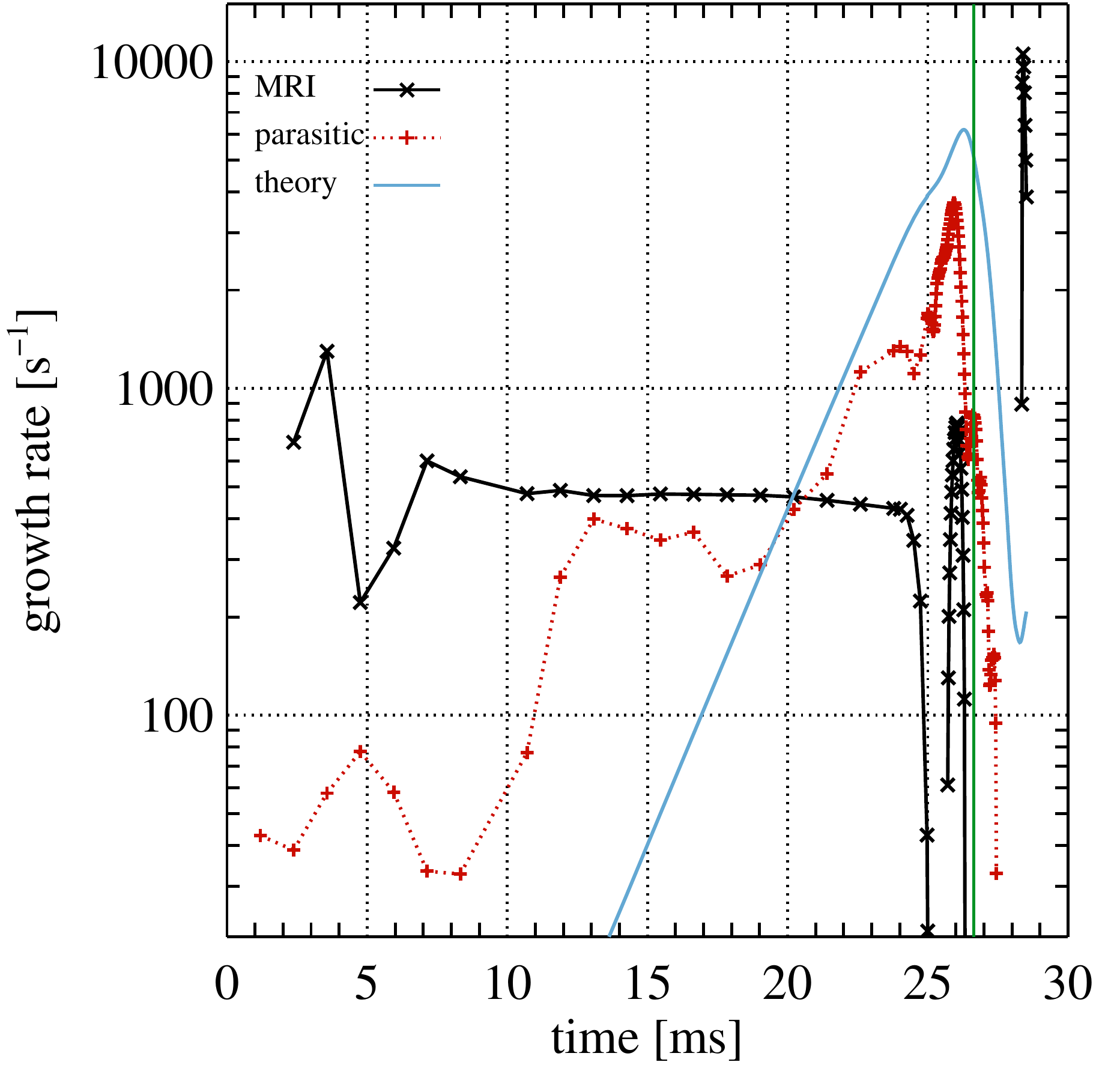}
\caption{Time evolution of the MRI and parasitic modes in simulation
  \#A8a. \emph{Upper left:} average magnetic energy density associated
  with the MRI channels ($e_{\rm MRI, \alpha}$) and the parasitic
  instabilities ($e_{\rm p, \alpha}$) for different components
  $b_\alpha$ of the magnetic field. In the inset, the phase around the
  MRI termination is presented. The vertical green line denotes
  termination time, $t = 26.7\, \ms$.  \emph{Bottom left:} MRI and
  parasitic growth rate calculated from the $b_r$ component.  The
  theoretically expected growth rate of the parasitic instabilities
  (Eq.~\ref{eq:gammakh}) is marked with the blue curve.  \emph{Upper
    right} and \emph{bottom right} panels are analogous to their left
  counterparts, but for the velocity components.  }
\label{fig:parasitic}
\end{figure*}

\begin{figure*}
\centering
\includegraphics[width=0.48\linewidth]{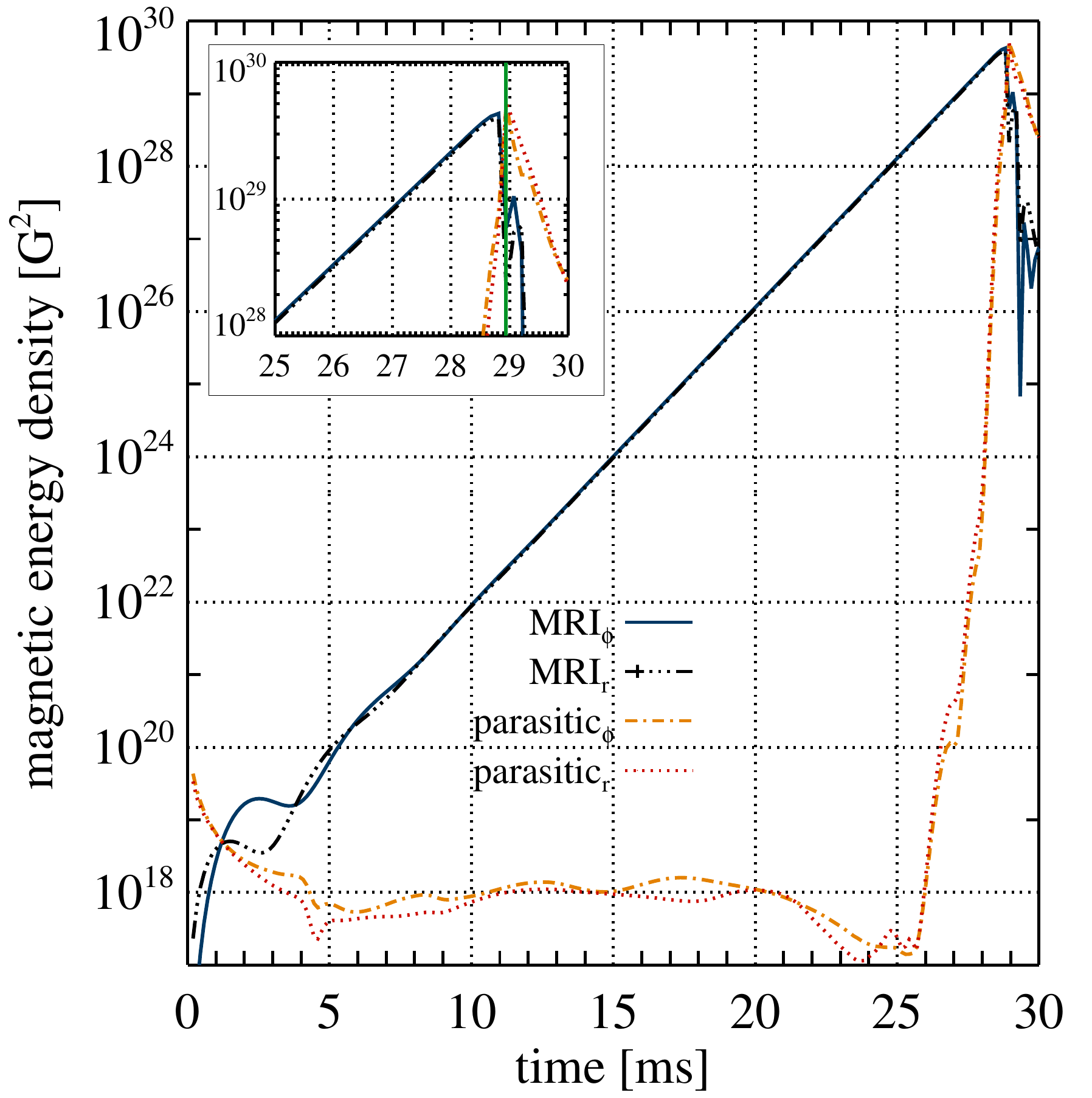}
\includegraphics[width=0.48\linewidth]{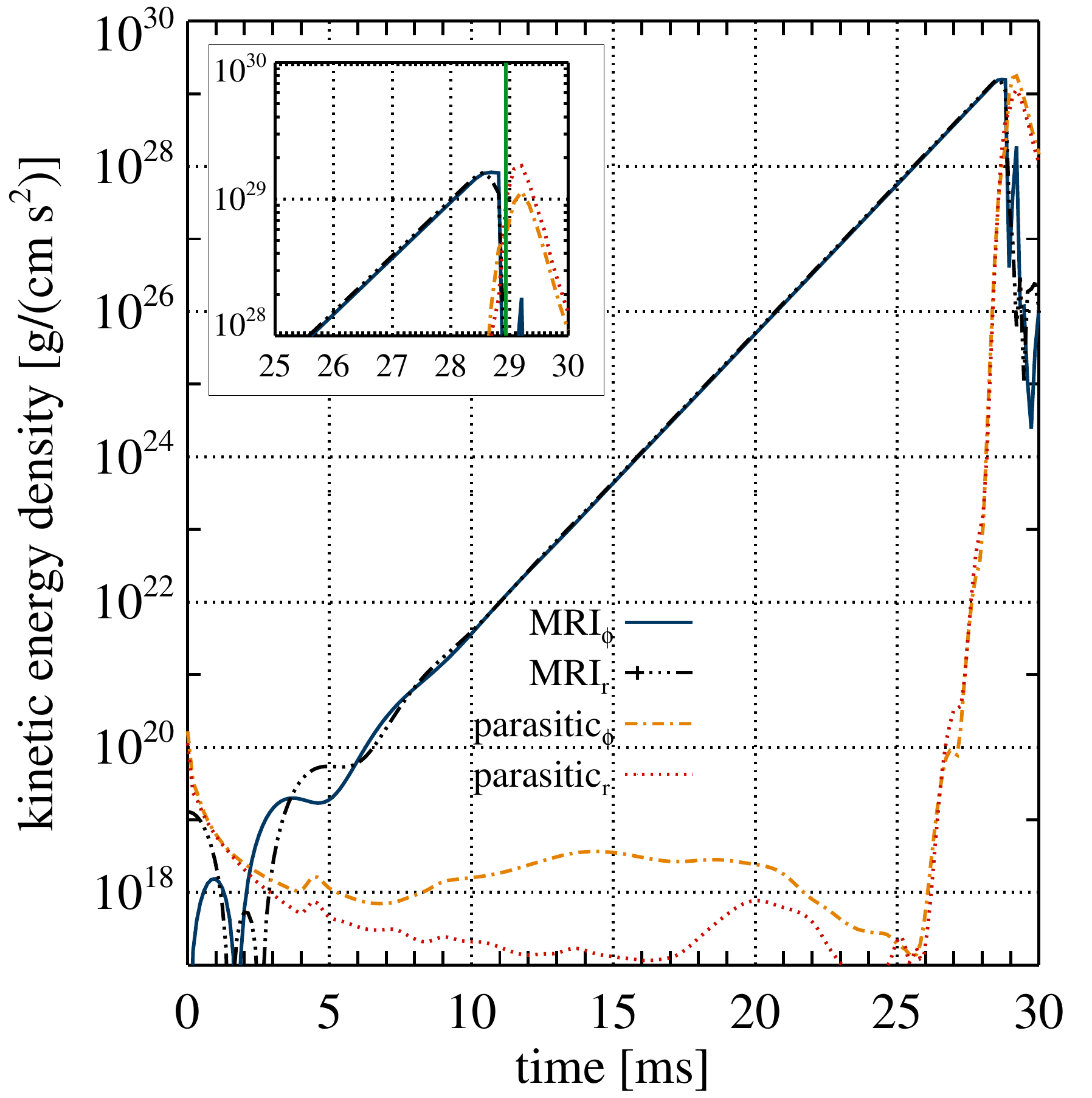}
\includegraphics[width=0.48\linewidth]{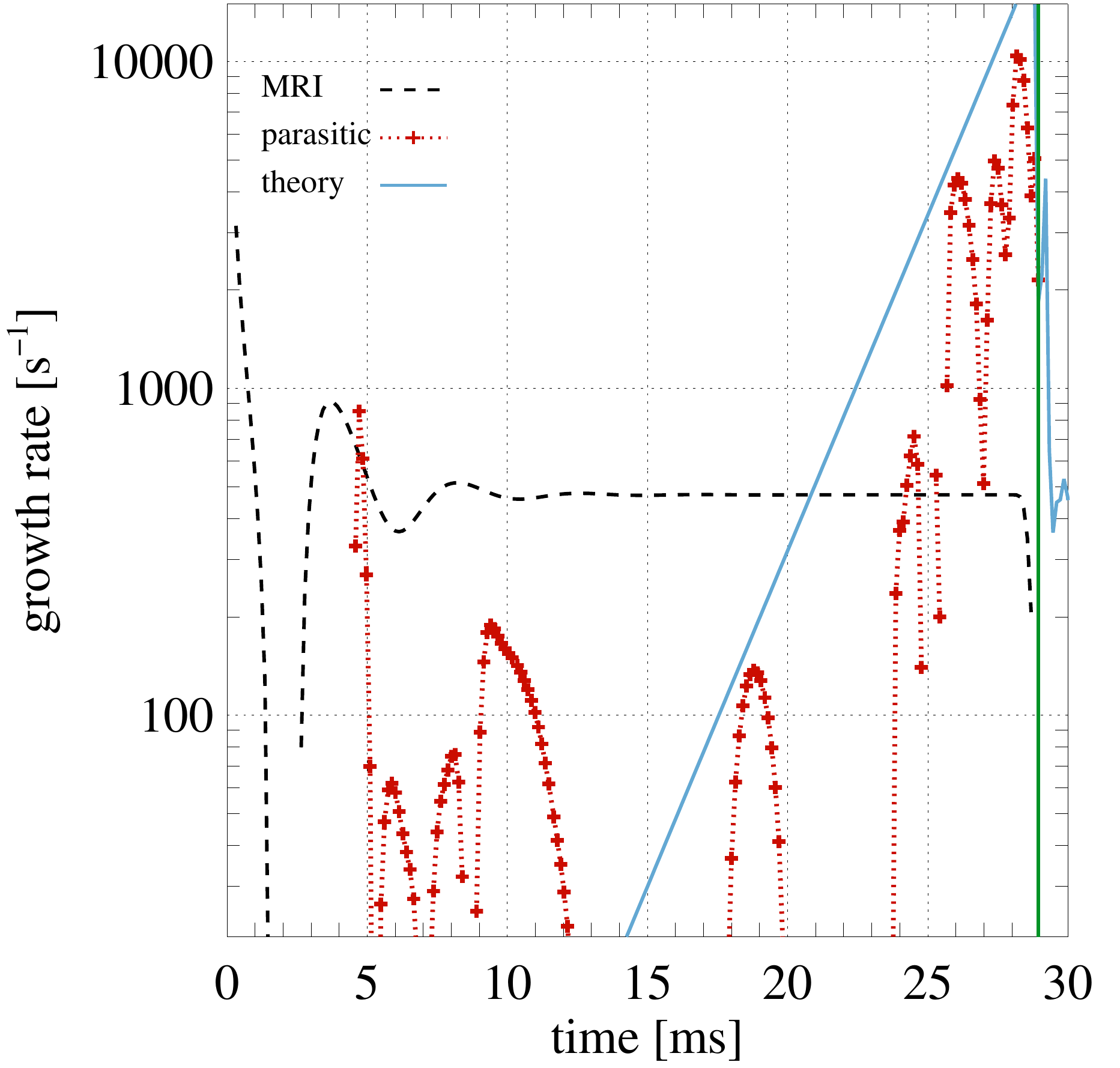}
\includegraphics[width=0.48\linewidth]{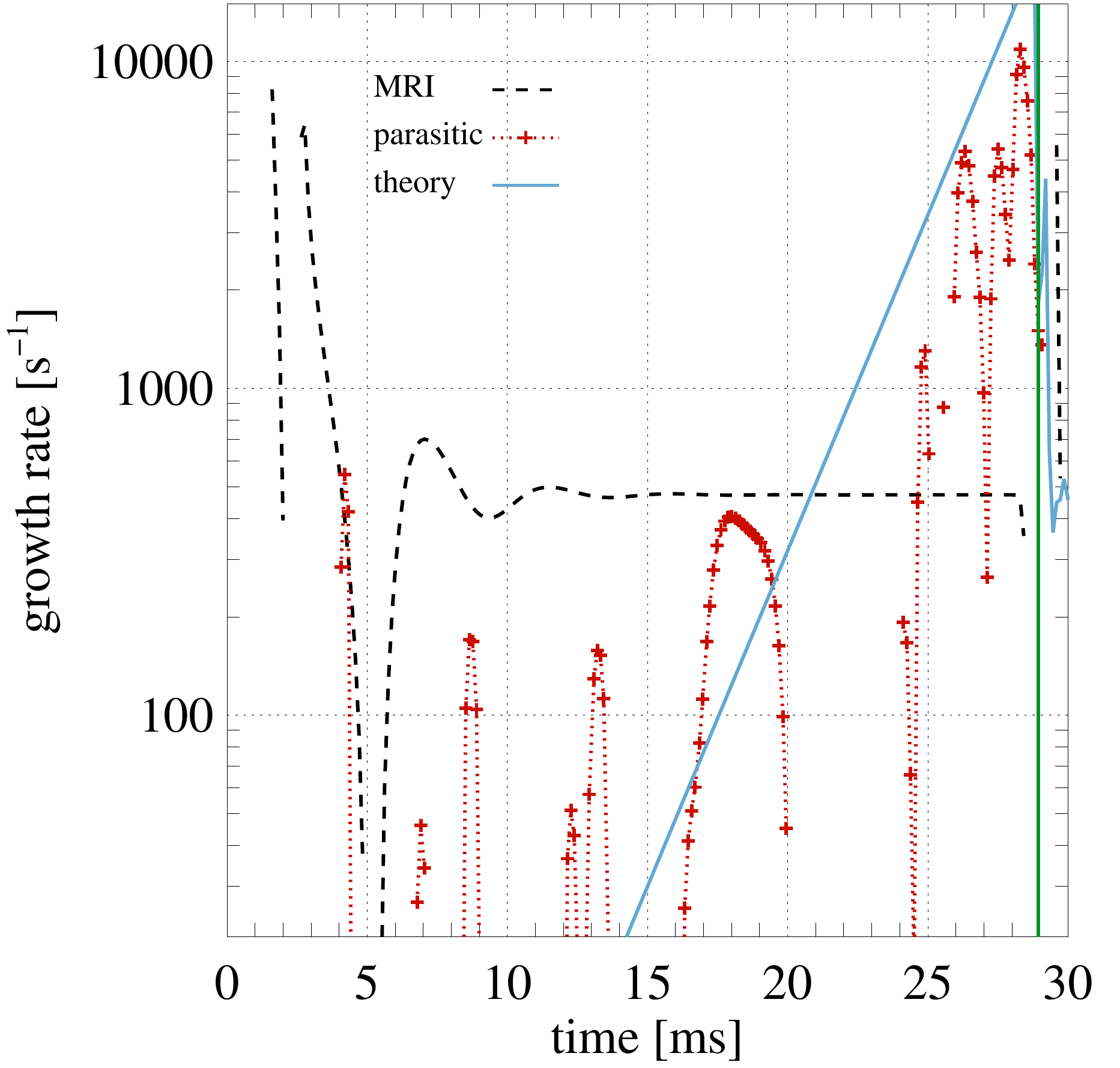}
\caption{Same as \figref{fig:parasitic}, but for \textsc{Snoopy} 
 simulation \#S19. The vertical green line  denotes termination 
 time $t = 28.8\, \ms$\trcross{$\sim 28.9\, \ms$}.
}
\label{fig:parasitic_Snoopy}
\end{figure*}

Figure\,\ref{fig:parasitic} (upper panels) shows the time evolution of
the magnetic and kinetic energy density for both the MRI and the
parasitic instabilities (given by
Eqs.~\ref{eq:mri_mag}--\ref{eq:p_kin}; see
\figref{fig:parasitic_Snoopy} for the corresponding results of a
\textsc{Snoopy} simulation).  The MRI (Fourier) modes grow
exponentially with time at a constant rate from $t \approx 8\,\ms$ to
$t \approx 24\,\ms$.  The average magnetic and kinetic energy density
$e^{\rm mag}_{{\rm p},\alpha}$ and $e^{\rm kin}_{{\rm p},\alpha}$,
respectively, of the parasites begin to grow super-exponentially at
$t \approx 20\, \ms$ from a value of about $6$ orders of magnitude
smaller than that of the MRI.  According to termination criterion I,
the MRI termination should occur already at $t \approx 21\, \ms$ (see
bottom panels of \figref{fig:parasitic}) however, we observe that the
MRI keeps growing for $\sim 5\, \ms$ 
\vlc{($\sim 2.7\,\gammamri^{-1}$)} longer
 \cross{. Indeed, the termination criterion II provides a better
   proxy for the actual termination of the MRI growth. As can be seen
   from the upper panels of} 
\trcross{the
   actual termination happens a bit before the amplitude of the
   parasitic modes equals that of the MRI ones, when the parasitic
   amplitudes become about half of the MRI amplitudes.  As we have
   mentioned before, termination criterion I should only be treated as
   an indicator that once it is fulfilled, the MRI termination will
   happen shortly (i.e.\, after a time of the order of a few MRI time
   scales) afterwards.}
 \vlc{until the amplitude of the parasitic modes becomes comparable to
   the amplitude of the MRI (upper panels of \figref{fig:parasitic}).
   Therefore, we conclude that termination criterion II provides a
   better proxy for the termination of the MRI growth.}

\cross{Having ruled out one of the proposed termination criteria, we
  will now investigate the other criterion in more detail.  The most
  striking feature of the end of the MRI is the maximum of $\MMM$
  (defined in }
\cross{ However, the total
  Maxwell stress may contain non-negligible contributions of the
  parasites and not only of the MRI channels and, thus, its maximum
  does not necessarily correspond to the maximum amplitude of the MRI
  channels.  Hence, it would be better to define the MRI termination
  as the point at which the amplitudes of the Fourier modes related to
  the MRI (Eq.\ } 
\cross{ reach their maximum.  The
  difference, however, turns out to be minor in \textsc{Aenus}
  simulations.  For simulation \#A8a, we were able to compare the two
  definitions.  We find comparable termination times of
  $t = 26.3\, \ms$ (maximum of channel Fourier modes) and
  $t = 26.7\, \ms$ (maximum of $\MMM$).  Consequently, termination is
  delayed w.r.t. the prediction of criterion I by a time of
  $\sim 2.7 \gammamri^{-1}$.  Note, however, the contamination
  introduced by our boundary conditions can somewhat reduce the time
  between when the termination criterion I is fulfilled and the MRI
  termination (see also Sec.\ } 
 \cross{ where we
  discuss an analogous simulation done with \textsc{Snoopy}, where
  this time is closer to $\approx 3 \gammamri^{-1}$). }

\vlc{We finally note that the time} 
\dvlc{interval between the
  fulfillment of the termination criterion I and MRI termination},
\vlc{$\sim 2.7 \gammamri^{-1}$, can be somewhat reduced by the
  contamination }\dvlc{caused by the}
\vlc{boundary conditions} \dvlc{used} \vlc{in \textsc{Aenus}
  simulations (see also Sec.\ \ref{sec:Snoopy_sim}, where for an
  analogous \textsc{Snoopy} simulation, this time} \dvlc{interval}
\vlc{is $\approx 3 \gammamri^{-1}$).  In
  Appendix \ref{app:aenus}, we discuss in more detail the influence of
  the boundary conditions used in \textsc{Aenus} simulations, as well
  as we make a more detailed comparison of \textsc{Aenus} and
  \textsc{Snoopy} simulations.}

\subsubsection{Snoopy simulation}
\label{sec:Snoopy_sim}

We repeat an analogous analysis of simulation \#S19 done with
\textsc{Snoopy}. Figure\,\ref{fig:parasitic_Snoopy} shows the time
evolution of the magnetic and kinetic energy density for both the MRI
and the parasitic instabilities (given by
Eqs.~\ref{eq:mri_mag}--\ref{eq:p_kin}).  The MRI (Fourier) modes grow
exponentially with time at a constant rate from $t \approx 8\,\ms$ to
$t \approx 27\,\ms$.
\cross{The amplitude of parasitic instabilities is constant in the
  first $25\,{\rm ms}$ and then experiences a fast growth.}
Like in the \textsc{Aenus} simulation, the parasitic instabilities
growth rate increases with the MRI amplitude in agreement with
theoretical expectations. An interesting feature that appears in the
\textsc{Snoopy} simulation is that the growth rate of the parasitic
modes shows oscillations with time with a period
$\tau \approx 2\tto 3 \, \ms$ (bottom panels of
Fig.\,\ref{fig:parasitic_Snoopy}).  These oscillations could be caused
by shear, as argued by \citet{Latter_et_al_2}.  A parasitic mode
wavevector has the optimal orientation only for a limited time before
its radial component becomes too large for an efficient growth.  Note
that according to \Eqref{eq:tau}, we would expect in this simulation
$\tau = 2.27 \, \ms$, which agrees very well with the observed period
of the oscillations.  This seems to hint that any future parasitic
model should take the influence of the shear into account, as its
influence on the parasitic modes is non-negligible.

\cross{The average magnetic and kinetic energy density
  $e^{\rm mag}_{{\rm p},\alpha}$ and $e^{\rm kin}_{{\rm p},\alpha}$,
  respectively, of the parasites begin to grow super-exponentially at
  $t \approx 25\, \ms$ from a value of about $10$ orders of magnitude
  smaller than that of the MRI.  The Maxwell stress $\MMM$ assumes its
  maximum at $t = 28.9\,\ms$, and then the MRI growth is terminated
  and the channels are disrupted.}

\cross{We can conclude that both termination definitions are
  consistent within $0.1\,{\rm ms}$ with the amplitude of the
  parasitic instabilities being equal to the MRI amplitude
  (termination criterion II) but are clearly inconsistent with
  criterion I.  Parasitic instabilities reach the growth rate of the
  MRI at $t \approx 23\, \ms$, but the termination happens only
  $\approx 6\, \ms \approx 3 \gammamri^{-1}$ later.}

\dvlc{The growth rate of parasitic instabilities reaches} \vlc{the
  growth rate of the MRI at $t \approx 23\, \ms$ (termination
  criterion I; bottom panels of \figref{fig:parasitic_Snoopy}).}
\dvlc{However, termination occurs only about $6\,\ms$
  ($\approx 3 \gammamri^{-1}$)} \vlc{later when the amplitude of the
  parasitic modes becomes comparable to} \dvlc{that} \vlc{of the MRI
  (termination criterion II; upper panels of
  \figref{fig:parasitic_Snoopy}).}  \dvlc{At $t \approx 25\, \ms$,
  both, the average magnetic energy density
  $e^{\rm mag}_{{\rm p},\alpha}$ and the kinetic energy density
  $e^{\rm kin}_{{\rm p},\alpha}$} \vlc{of the parasites begin to grow
  super-exponentially} 
\vlc{(until MRI termination at $t = 28.9\,\ms$) from} \vlc{a value of about $10$ orders of
  magnitude} \dvlc{smaller than} \vlc{the amplitude of the MRI.}

\subsection{Amplification factor}

\subsubsection{Dependence on the initial perturbations in Aenus}

\vlc{As we discuss in more detail in Appendix \ref{app:aenus}, the}
boundary conditions used in simulations performed with \textsc{Aenus}
do affect the exact value of the amplification factor and its
dependence on the initial parasitic perturbations.  Intuitively, one
would expect (somewhat) larger amplification factor for smaller
initial parasitic perturbations (see also Eqs.~\ref{eq:AAA_ln} and
\ref{eq:AAA_Lat}).  However, independently of the initial
perturbations, non-axisymmetric perturbations introduced by our radial
boundary conditions can be used by parasitic instabilities as their
seed perturbations.  These can be seen when comparing the
amplification factor of simulations \#A8a, \#A8b, \#A8c, which are
done with the same initial conditions (physical and numerical) but
different realisations of random initial perturbations with the same
normalisation $\delta$.  However, the amplification factor is
basically the same in all three simulations, i.e.  $\AAA \simeq 21$.

This fact can be seen even more clearly when comparing simulations
\#A9a and \#A9b. In the former, we used standard initial
perturbations, whereas in the latter we put $k_z = 3 \kmri$ (see
Eq.~\ref{eq:v_with_sin}), i.e.\ we perturbed the system with a mode
that should be stable against MRI \citep[cf. PC08,][for the
instability criterion]{Rembiasz_et_al}.  The only difference between
these two simulations was, that in the latter an MRI channel was
formed a few milliseconds later and MRI was terminated $\sim 10\,\ms$
later.  However, the amplification factor was basically not affected
($\AAA = 21.0$ and $\AAA = 20.0$, respectively). A more detailed
discussion of this simulation can be found in \cite{Rembiasz}.

In spite of the boundary conditions used in \textsc{Aenus}
simulations, we could say that because of the fact that due to these
artificial non-axisymmetric perturbations, the amplification factor
does not have any random scatter, this has the upside that we could
perform convergence studies. On the other hand, in order to reliably
assess the dependence of the amplification factor on the initial
perturbations, we have validated the \textsc{Aenus} results by
employing a completely different code, \textsc{Snoopy}, where the
boundary conditions are much easy to handle (though under the
restrictions of applicability spelled out in Sect.~\ref{sec:Snoopy}).

\subsubsection{Dependence on the initial perturbations in Snoopy}
\label{sec:sususeSn}

\begin{figure}
\centering
\includegraphics[width=1\linewidth]{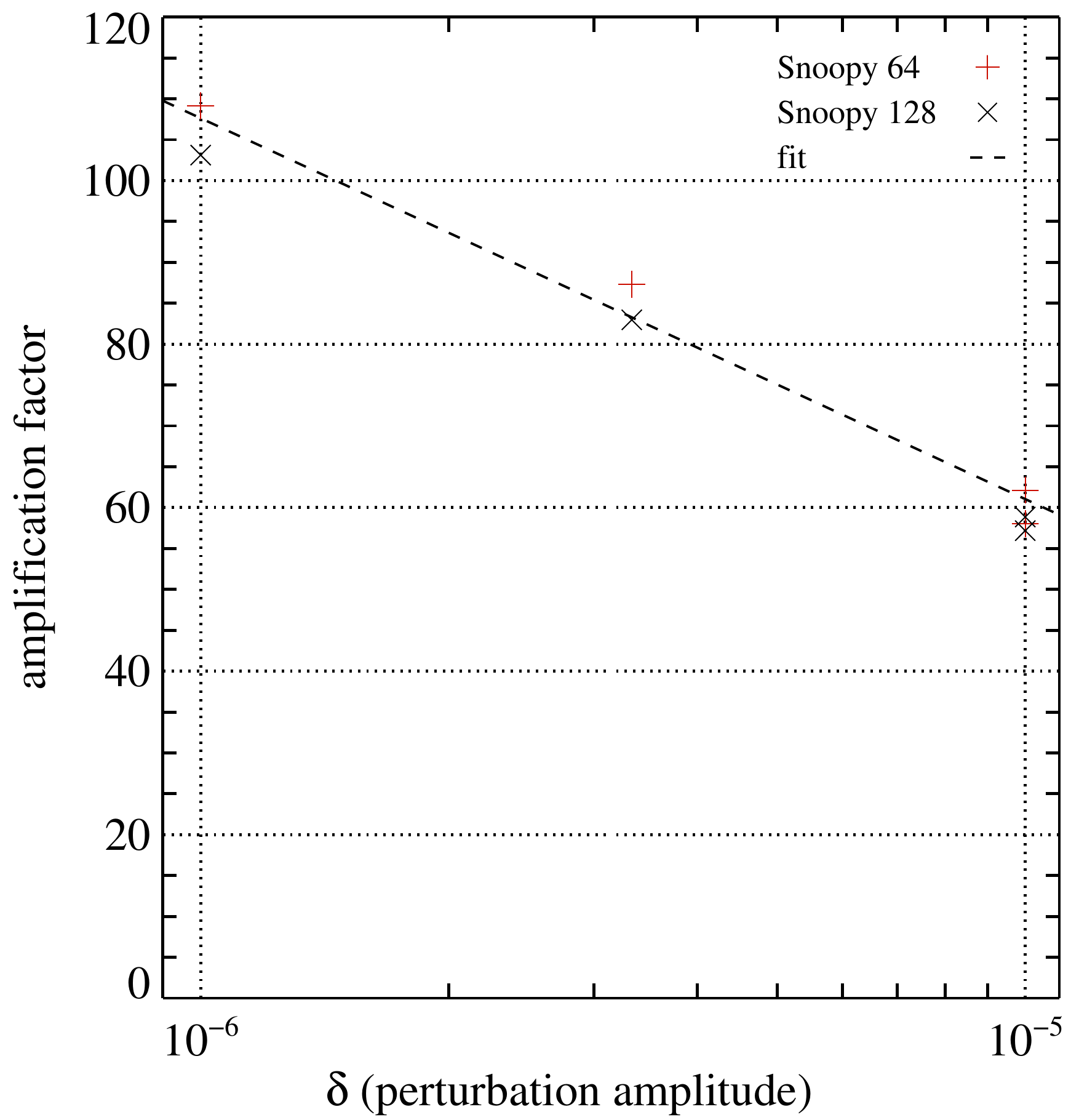}
\includegraphics[width=1\linewidth]{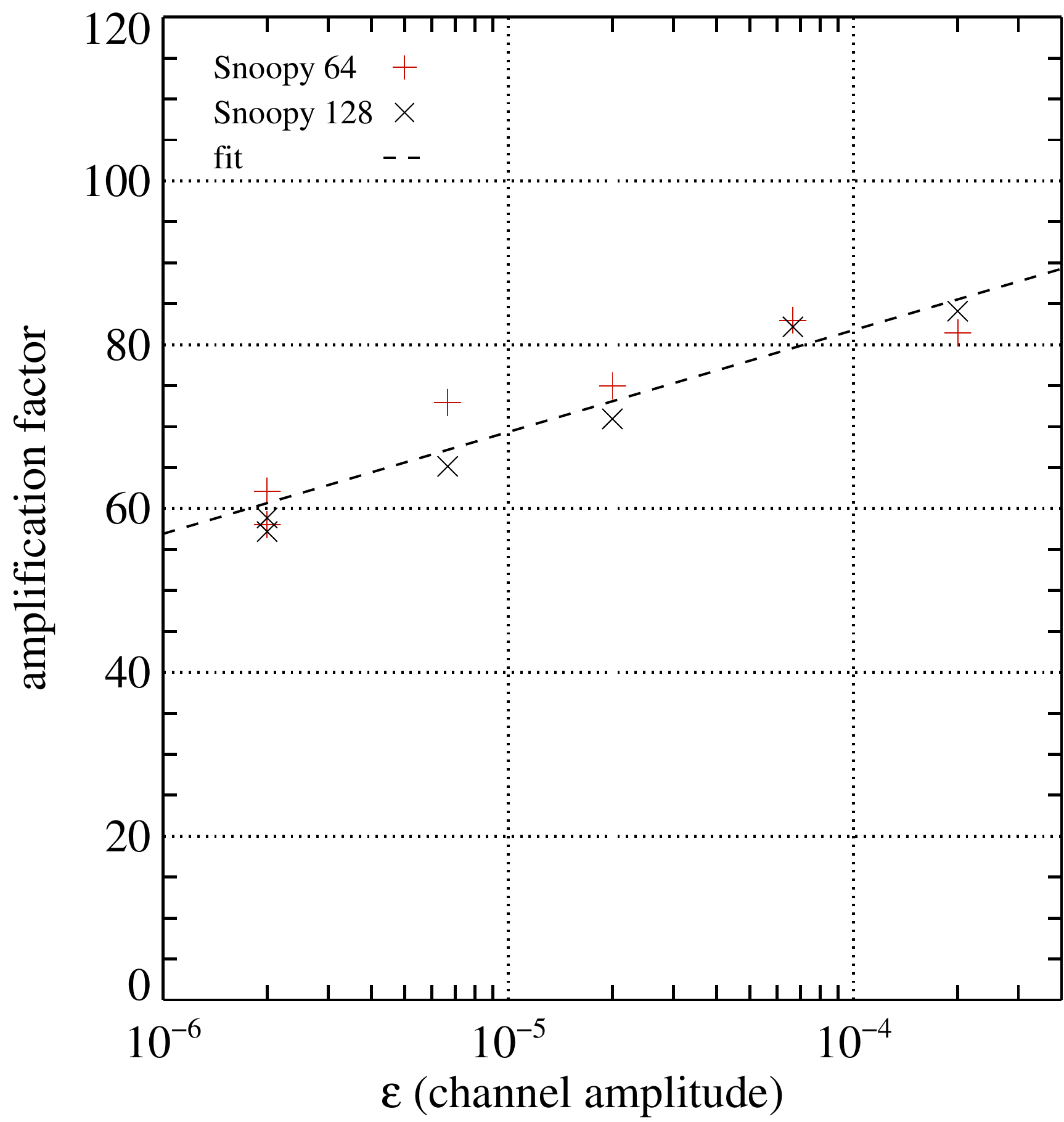}
\caption{ \emph{Top:} amplification factor (defined in
  Eq.~\ref{eq:AAA}) dependence on the initial parasitic perturbations
  (proportional to $\delta$, see Eq.~\ref{eq:v_with_sin}) for
  simulations \#S15a\tto \#S15d (\emph{red plus} symbols, 64 zones per
  MRI channel) and \#S16a\tto \#S16d (\emph{black crosses}, 128 zones
  per MRI channel) done with \textsc{Snoopy}. The fit to the data
  points was done according to \Eqref{eq:adc}.  \emph{Bottom:}
  amplification factor dependence on the initial channel perturbations
  (proportional to $\epsilon$) for simulations \#S15c\tto \#S15h and
  \#S16c\tto \#S16h.  }
\label{fig:chan_perts_sims}
\end{figure}

To test the dependence on the amplitude of the initial random
perturbations, which are mainly a seed for KH parasitic instabilities,
i.e.\ parameter $\delta$ in \Eqref{eq:v_with_sin}, we ran simulations
\#S15a\tto \#S15d and \#S16a\tto \#S16d with the same initial magnetic
field strength $b_{0z} = 1.18\times 10^{13}\,\mathrm{G}$ in all
models, but with two different resolutions and varying amplitudes of
the initial perturbations in the range
$\delta=(0.1\tto 1) \times 10^{-5}$ (upper panel of
\figref{fig:chan_perts_sims}).  We note that the models \#S15c and
\#S15d have the same initialisation but, because of the stochastic
nature of the initial perturbations they are run to assess the scatter
of the final results as a function of the initial randomness imposed
by the perturbations parametrised with the amplitude $\delta$. The
same comment applies to models \#S16c and \#S16d.  The amplification
factor is in the range $\AAA \simeq 60\tto110$, which is by a factor
$\sim 3\tto 5$ larger than in \textsc{Aenus} simulations.

In order to test the influence of the initial channel amplitude on the
amplification factor, we consider the sets of models \#S15c\tto \#S15h
and \#S16c\tto \#S16h. Both sets differ in resolution and within each
set we vary the initial channel amplitude in the range
$\epsilon=(0.2\tto 20) \times 10^{-5}$ (bottom panel of
\figref{fig:chan_perts_sims}).  We see a logarithmic dependence of the
amplification factor on the amplitudes of both the initial
perturbations and of the channel modes.
To these simulation results we fitted a function
\begin{equation}
  \label{eq:adc}
  \AAA( \epsilon, \delta) = a \ln \epsilon + d \ln \delta + C,
\end{equation}
obtaining $a = 5.4 \pm 0.55$, $d = -20.2 \pm 1.2$, and
$C = -101 \pm 13$.  Now, we can compare these results with two
different theoretical predictions for the amplification factor given
by \eqsref{eq:AAA_ln} and (\ref{eq:AAA_Lat}).
If we assume that $\tildevkh \propto \delta$, $\sigma = 0.27$, and
$\ln \AAA \approx \text{const.}$ in this parameter range, from
\Eqref{eq:AAA_ln}, we would theoretically expect $ d = -1.85$.  We
obtain a discrepancy of one order of magnitude.  When calculating the
parasitic growth rate (Eq.~\ref{eq:gammakh}), PG09 neglected the
influence of the background shear (and of the Coriolis force) which,
as pointed out by \citet{Latter_et_al_2} can lead to a reduction of
the growth rate.
Taking this effect into account in an approximate way,
\Eqref{eq:AAA_Lat} predicts $ d = -3.41 $.  Moreover, taking into
account the growth of the magnetic field in the time interval $\tau$
by a factor of $\approx 3$ (see the discussion below Eq.\
\ref{eq:AAA_Lat}), would lead to $ d = -10.1 $, which is closer to the
simulation results.  This discrepancy could be caused by the
extrapolation of \Eqref{eq:gammakh} to the non-linear regime of the KH
instability, where we, however, expect a growth rate reduction.
Taking all these effects into account, we could expect a significant
growth rate reduction. Note that from the fit we could estimate that
the {\em effective} value of $\sigma$ is $\sigma_{\rm eff} = 0.0125$,
i.e.\ the KH growth rate should be one order of magnitude lower than
predicted by \Eqref{eq:gammakh}.  However, we should recall that
$\sigma$ cannot really be a constant, since by definition, after
saturation the parasitic modes should not grow any longer, implying
$\sigma \rightarrow 0$. Very likely, the proportionality between
$\gamma$ and $\vc$ expressed in Eq.\, (\ref{eq:gammakh}) is a
linearisation of a more general relation of the form
$ \gammakh = \kmri f(\vc)$, where $f(\vc)$ is a non-linear function of
the velocity of the MRI channels. However, exploring modifications of
the parasitic model theory that include such kind of non-linearities
is beyond the scope of this paper.

To study the dependence of the amplification factor on the initial
channel amplitude, if we assume that $\tildevc \propto \epsilon$,
$\sigma = 0.27$, and again $\ln \AAA \approx \text{const.}$ from
Eqs. (\ref{eq:AAA_ln}) and (\ref{eq:AAA_Lat}) , we would theoretically
expect $a= 0$, i.e.\,the amplification factor should be independent of
$\epsilon$ (note that we neglected the last term in the RHS of
\Eqref{eq:AAA_ln}, as $ \frac{\tildevc}{\caz} \ll 1$ in our
simulations).  Thus, we conclude that present predictions within the
parasitic instability model fail to predict this dependence correctly.

To test the dependence of the amplification factor not only on the
amplitude of the initial perturbations but also their form, we ran
simulations \#SA15a-e obtaining $\AAA \approx 90$, which is
somewhat higher than for the corresponding simulations \#S15c--d for
which $\AAA \approx 60$. Taking into account that the MRI termination
is a highly non-linear process, we conclude that there is some, but
not very strong dependence.

Finally, it is worthwhile comparing predictions given by two different
numerical codes for similar initial conditions.  In the best resolved
\textsc{Aenus} simulations with
$b_{0z} = 1.22\times 10^{13}\,\mathrm{G}$, i.e.\, models \#A10 and
\#A11, we obtained $\AAA \approx 19$, which is a considerably smaller
value than in analogous \textsc{Snoopy} simulations, i.e.\, models
\#S15a--e (with $b_{0z} = 1.22\times 10^{13}\,\mathrm{G}$) in which
$\AAA \approx 90$.  This difference can be probably attributed to the
following \vlc{fact. The} imperfect radial boundary conditions
continuously pollute the numerical data on the grid (see the bottom
panel of \figref{fig:04}).  These perturbations of both physical and
numerical origin can act as seed perturbations for the parasitic
instabilities. For \textsc{Aenus}, they seem to dominate the initial
random perturbations of amplitude $\delta$, which therefore are
completely insignificant in comparison. Consequently, the termination
of the MRI is decoupled from the parameter $\delta$.  Comparing the
upper panels of Figures \ref{fig:parasitic} and
\ref{fig:parasitic_Snoopy}, we see that in the \textsc{Aenus}
simulations, the non-axisymmetric perturbations grow from $5$ to
$20\, \ms$, whereas they remain roughly constant in the
\textsc{Snoopy} simulations. \cross{At $t \approx 20\, \ms$} 
\vlc{In model \#A8a}, the
non-axisymmetric perturbations \cross{in model \#A8a} start to grow
super-exponentially with time \vlc{from $t \approx 20\, \ms$}\cross{,
  which}\vlc{. This} can be
attributed to the genuine parasitic instabilities.  Therefore, we
could equally well start this simulation at $t \approx 20\, \ms$ with
MRI channels determined by $\epsilon \approx 10^{-2}$ and random
perturbations of roughly $ \delta \approx 10^{-3}$.
%
%
For these parameters, \Eqref{eq:adc} yields an amplification
  factor $\AAA(\textsc{Aenus}) = 14 \pm 24$, \vlc{ i.e.\, smaller than
    $38$}, which is compatible with the value measured in the
  simulation, i.e.\ $\AAA = 18.6$ (for the amplitudes used in the
  actual simulation, i.e.\, $\delta = 10^{-5}$ and
  $\epsilon = 2 \times 10^{-6}$, the fitting formula predicts
  $\AAA = 61 \pm 34$, \vlc{ i.e.\, $\AAA > 27$} ).  This result allows
  us to conclude that, if we account for the specific way the
  parasites are seeded, both sets of simulations can be understood by
  a common approximate theory, which we, thus, regard as a reasonable
  description of the physics.

To test this hypothesis, we ran additional \textsc{Snoopy} simulations
\#SCA15a--g, \#SCA15a--g, and \#SCR15a--c with the same magnetic field
strength, but much higher initial amplitudes (proportional to
$\epsilon$) of both full MRI channels (and not only perturbations in
the $v_r$ component which facilitated the development of the channels)
and random perturbations (proportional to $\delta$).  In all these
simulations we set $\epsilon = 1.2 \times 10^{-2}$ and
$\delta = (0.02 \tto 2.27) \times 10^{-2} $, which should mimic the
conditions in the \textsc{Aenus} simulation \#A8a at
$t \approx 21\, \ms$. We measured in these additional \textsc{Snoopy}
simulations that the amplification factor indeed decreased by a factor
of $2\tto 3$ to $\AAA \approx 18 \tto 50$.  This confirms our
hypothesis that some part of the discrepancy between \textsc{Aenus}
and \textsc{Snoopy} simulations can be explained by spurious
perturbations induced by the imperfect radial boundary conditions used
in the former code.

We note that the differences (by a factor of 5) in the amplification
factor computed from \textsc{Aenus} and \textsc{Snoopy} probably
result from differences in \vlc{the radial boundary conditions,} the
  physical assumptions, \cross{(compressibility)} and \cross{in} the
 numerical schemes employed \cross{by} \dvlc{in} both
  codes.
\cross{The MRI termination is a highly non-linear process developing
  over a very short time scale. Hence even tiny differences in the
  growth rates of the parasitic instabilities can lead to
  non-negligible differences of the final MRI termination amplitude
  and hence the amplification factor.}

\subsubsection{Dependence on the initial magnetic field strength 
                      in Snoopy}

Finally, we want to test the most relevant prediction of the parasitic
model, i.e.\ the dependence of the amplification factor on the initial
magnetic field strength.  As in our simulations we used the most
optimistic initial values for the magnetic field strength
\citep[i.e.\, $b_{0z} \approx
10^{13}\,\mathrm{G}$;][]{Obergaulinger_2014}, it is of crucial
importance to know whether in CCNSe with weaker initial magnetic
fields the MRI will amplify the initial fields to the same value or by
the same factor (or something in between).  According to the estimate
of PG09, the MRI should amplify the initial magnetic field by a
constant factor (Eq.~\ref{eq:a1.9}).  However, their estimate was done
using termination criterion I (Eq.~\ref{eq:term1}), which as we have
shown with our numerical models, is not an exact
predictor of the MRI termination.  According to our estimates done
within the parasitic model using termination criterion II
(Eq.~\ref{eq:AAA_ln}), which as we have shown yields a much better
prediction of the MRI termination, the amplification factor should be
independent of the initial magnetic field strength provided that the
ratio $\tildevkh / \caz$ is constant.  Similarly, the estimate of
$\AAA$  that we can infer from \citet{Latter_et_al_2}
(Eq.\,\ref{eq:AAA_Lat}) also predicts no dependence on the initial
magnetic field strength.  Therefore, in order to address whether
$\AAA$ is independent from the initial magnetic field strength, we ran
several numerical models.

\begin{figure}
\centering
\includegraphics[width=1\linewidth]{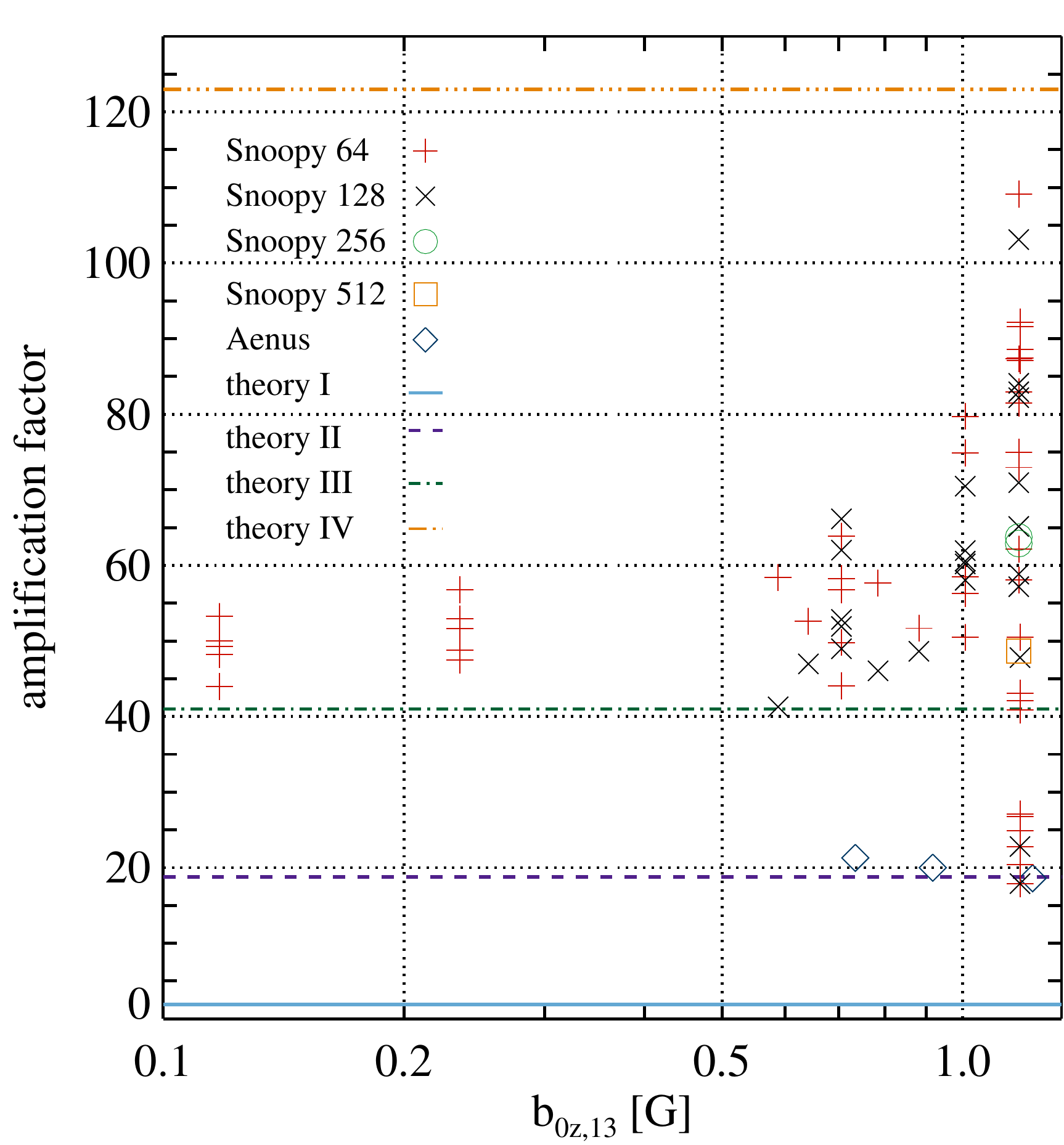}
\caption{Amplification factor as a function of initial magnetic field
  in the best resolved simulations done with \textsc{Aenus}
  (i.e.~models \#A4 \#A6 and \#A11 from \tabref{tab:aenus}) and all
  simulations done with \textsc{Snoopy} (\tabref{tab:Snoopy}).
With light blue solid, purple dashed, green dash dot, and orange dash
dot dot lines,  we  respectively marked theoretical estimates I
 (Eq.~\ref{eq:a1.9}),  II  (Eq.~\ref{eq:a19}), III (Eq.~\ref{eq:a41}),
and   IV (Eq.~\ref{eq:a123})
of the amplification factor.
Note that the results shown for \textsc{Snoopy} for the same initial
magnetic field strength and resolution display a relatively large
scattering as a result of the different initial amplitude of the
perturbations.}
\label{fig:amp_fac}
\end{figure}

In \figref{fig:amp_fac}, we present all our best resolved models done
with \textsc{Aenus}, i.e.\ \#A4 \#A6 and \#A11 with the initial
magnetic field strength
$b_{0z} = (0.73\tto 1.22)\times 10^{13} \,\mathrm{G}$, and all
\textsc{Snoopy} simulations with
$b_{0z} = (0.12\tto 1.18) \times 10^{13} \,\mathrm{G}$.  Because of
the scatter of the amplification factor in \textsc{Snoopy} simulations,
it was impossible to perform proper convergence tests (we would need
many more simulations to compute proper averages).  We ran only a few
simulations with a very high resolution, i.e. 256 and 512 zones per
MRI channel (models \#S17a,b and \#S18, respectively), which did not
differ from those ran with lower resolutions. Therefore, we conclude
that simulations done with 64 and 128 zones per MRI channels should
give reasonably good predictions.  We also marked in
\figref{fig:amp_fac} the amplification factor estimated by PG09
(Eq.~\ref{eq:a1.9}; light blue solid line) and us (dark blue dashed
line) in \Eqref{eq:AAA_ln} assuming initial parasitic amplitudes
$\tildevkh/\caz \approx 10^3$ (a representative value for our
simulations).  When looking at all our simulation results we can
conclude that \textsc{Aenus} simulations give a value of $\AAA$ lower
by a factor of $1\tto 6$ than that obtained with \textsc{Snoopy}. The
discrepancy can be partly attributed to the
 boundary condition
\cross{and the influence of compressibility}  that we discussed before. 

Beyond the code agreement on the exact value of $\AAA$, the
amplification factor seems to be indeed independent of the initial
magnetic field strength. This conclusion can be drawn on the light of
the results of simulations \#S1a-- \#S1e with an initial magnetic
field lower by one order of magnitude than the rest of the models,
i.e.\,$b_{0z} = 0.12 \times 10^{13} \,\text{G}$. Even in these cases,
the amplification factor stays the same within the random scatter,
i.e.\,$\AAA \approx 50$ (Fig.\,\ref{fig:amp_fac}).

Note that this result is also consistent with the results of
\citet{Obergaulinger_et_al_2009}, who found a numerical scaling
$\MMMterm \propto b_{0z}^{16/7}$, which would translate to
$\AAA \propto b_{0z}^{0.14}$ (i.e.\, a very weak dependence of the
amplification factor on the initial magnetic field strength). Note,
however, that as \citet{Obergaulinger_et_al_2009} did not perform
proper convergence studies, it is impossible to conclude based on
their results whether this scaling law was of a physical or of a
numerical origin. Indeed, \citet{Rembiasz_et_al} and
\citet{Rembiasz_et_al_ASTRONUM} observed dependence of $\MMMterm$ on
the numerical schemes and resolutions used in their studies.

\section{Summary}
\label{sec:summary}

Since the direct numeral simulations of
\citet{Obergaulinger_et_al_2009} confirming the theoretical
predictions of \citet{Akiyama_etal__2003__ApJ__MRI_SN}, there is no
doubt that the MRI can amplify the initial magnetic field (close to the
surface of a PNS) on a sufficiently short time scale in CCSNe produced
by rapidly rotating progenitors.  However, the limit of such a
magnetic field amplification during a CCSN explosion remained unknown.
In this paper, we aimed to address an important aspect of this
question, i.e.\, the factor by which the seed field of the core is
amplified during the exponential growth phase of the MRI.  To this
end, we performed shearing disc simulations with an Eulerian MHD code,
\textsc{Aenus}, solving full MHD equations and shearing box
simulations with a pseudo-spectral code, \textsc{Snoopy}, in the
incompressible approximation.  We compared our results to the
predictions of the parasitic model proposed and developed by GX94,
\cite{Latter_et_al}, PG09 and \citet{Pessah}.  Part of the predictions
of the parasitic model, i.e.\, that given CCSN conditions, the MRI
should be terminated by parasitic KH instabilities was confirmed in
direct numerical simulations of \citet{Rembiasz_et_al}.  However,
those authors neither performed systematical studies of the
amplification factor dependence on the initial conditions, nor
compared the field amplification obtained in their simulations to the
predictions of the parasitic model.

Within the parasitic-termination model, the MRI channel modes are
susceptible to secondary instabilities, in our case of KH type.  The
former grow at a constant rate given by the rotational profile of the
core, whereas the growth rate of the former is a function of the
amplitude of the MRI and, thus, increases continuously.  At some
point, they will be sufficiently strong to disrupt the MRI channel
modes and thereby terminate the MRI growth.  Based on this
observation, PG09 proposed two different criteria to identify the
moment of parasitic termination in their analytic model: termination
occurs when the growth rate of (initially developing much more slowly)
parasitic instabilities starts to exceed the growth rate of MRI
(termination criterion I), or when the amplitudes of the parasitic
instabilities reach the amplitudes of the MRI channels (termination
criterion II).

We tested these two termination criteria and found that in simulations
done with both codes, termination criterion II represents a better
description of the results. MRI termination occurs when parasitic
instabilities roughly reach the amplitudes of the MRI channels, which
happens roughly $3 \gammamri^{-1}$ after termination criterion I is
met.

Next, we compared the theoretical predictions of \citet{Pessah},
\citet{Latter_et_al_2} and our estimates based on \citet{Pessah} for
the amplification factor with our simulation results.  We find an
order-of-magnitude agreement of Latter's model with the numerical
simulations, although this model fails to explain all dependencies
accurately.  This better agreement could be due to the approximate
inclusion of the background shear by \citet{Latter_et_al_2}.  However,
differences could also be due to non-linearities at termination, not
considered in any theoretical estimate. From our numerical results we
cannot favour any of those possibilities.  Nevertheless, a more
elaborate description of the parasitic instabilities in the presence
of shear may be needed for an accurate prediction of the termination
amplitude.

Another prediction of these simplified models is that the
amplification factor should be independent of the initial magnetic
field strength. We tested this hypothesis with numerical simulations.
Our main finding of this paper is that independently of the initial
magnetic field strength, the MRI channel modes can amplify the seed
magnetic field by a factor of 20 to 100.  Once these magnetic field
values are reached, further MRI-driven magnetic field amplification is
halted as the MRI channels are attacked and destroyed by parasitic KH
instabilities.

It is true that, in principle, one could obtain an arbitrary value of
the amplification factor from our scaling relation (Eq.\ \ref{eq:adc})
by tuning the amplitudes of the initial perturbations.  However, for
physically plausible conditions found at the surface of the hot
proto-neutron star, the initial amplitudes are more likely to be of
the order $\sim 0.1 \tto 1$ of the rotational velocity. Under these
conditions, realistically expected values of the amplification factor
are of the order of $10$. In any case, even for a highly unperturbed
flow with perturbations $\sim 10^{-5}$ we could not find amplification
factors exceeding $\sim 100$.
\vlc{We stress that our results have been obtained under the most
  favourable conditions
for the MRI to develop concerning location and geometry of our
  computational boxes.
The fact that
the boxes
are located in the equatorial plane of the PNS optimises the
  topology of the magnetic field for the fastest possible growth
  \citep[see, e.g.\,][]{Balbus_Hawley__1998__RMP__MRI}.  
If even under
these {\em optimal} conditions the ability of the MRI channels as
  an agent to amplify the initial field is limited, we shall expect
  that anywhere else in the PNS the MRI channel-mode amplification be
  even smaller.}

\vlc{While our simulations allow us to draw conclusions on the field
  amplification by MRI channel modes, their applicability beyond the
  linear, local MRI modes needs to be assessed critically:}

\begin{itemize}
\item \vlc{The MRI might take other forms than that of channels (e.g.\,
    global modes) under realistic conditions especially for slower
    rotation rates.}
\item \vlc{Even if non-local modes can be excluded, the channels might
    not appear at all due to the presence of other instabilities.  As
    we have shown here, strong perturbations may quench the MRI
    channel modes very early, hence reinforcing our conclusion that
    the amplification is limited.}
\item \vlc{The long-lasting
  turbulent phase after the disruption of channels might strongly
    modify the amplification achieved by the channel modes because an
    MRI-driven dynamo could
be at work during this phase.}
\end{itemize}

The main conclusion that should be drawn by the supernova community is
that the amount of field amplification the MRI \vlc{channels} might
provide in CCSNe might be fairly limited.  This finding casts doubt on
the common procedure of starting global simulations with artificially
enhanced field strengths that is based on the expectation that the MRI
will provide rapid amplification of even very weak initial fields up
to equipartition levels with the kinetic (rotational) energy. 
\vlc{This may be true if an MRI-driven turbulent dynamo develops
  sufficiently fast and efficiently.}

\vlc{Beyond the implications for CCSNe, we also point out that our
  results are relevant for the standard setup assumed in models of
  progenitors of gamma-ray bursts. In those models a fairly large
  magnetic field ($\gtrsim 10^{15}\,$G) is needed to extract an
  ultrarelativistic outflow fed by the rotational energy of the
  central compact object. Regardless of whether such an object is a
  proto-magnetar or a black hole,}
\vlc{our findings suggest that the progenitor star should be endowed
   with an uncommonly large magnetic
  field prior to core collapse.
  Alternatively, a large-scale turbulent dynamo, driven either
by rapid rotation and convection, or by
 the MRI, might be able to operate on longer timescales
  \citep{Thompson_Duncan__1993__ApJ__NS-dynamo}.  However, it still
  remains to be proven that this is the case in global simulations,
  starting from realistic progenitors. This has been explored recently
  by \cite{Mosta_2015}, albeit in a regime in which magnetic field is
  already of magnetar strength.  }

\vlc{Additionally}, one should bear in mind the limitations of our
studies in which we neglected the influence of neutrinos
\citep[see][]{Guilet_et_al_2015}, buoyancy and entropy gradients
\citep[see][]{Obergaulinger_et_al_2009,Guilet_Mueller_2015}.
Nevertheless, we do not think that including these effects could lead
to much stronger magnetic field amplification by the MRI channel
modes.  Moreover, we did so far not consider any possible additional
field amplification by a MRI-driven dynamo acting in the turbulent
saturated state after the end of MRI channel mode growth.  However its
influence on the MRI termination as well as the influence of
compressibility remain unknown.

\appendix

\section{Rescaling of the simulations of Rembiasz et al.~(2016a)} 
\label{app:conversion}

\citet{Rembiasz_et_al}, following \citet {Obergaulinger_et_al_2009},
performed almost ideal MHD (i.e. with $\Ree, \Rm \ge 100$) simulations
in a computational domain centred around the equatorial plane at a
radius $\tilde{r}_0 = 15.5\,\km$, i.e.  in middle of a nascent PNS of
radius $r_{\mathrm{PNS}} \approx 30\,\km$.  However, according to the
recent estimates of \cite{Guilet_et_al_2015}, at these distances the
neutrino viscosity cannot be neglected and it can suppress the MRI.
Therefore, since the models can be suitably rescaled, we shifted the
centre of their computational domain close to the surface of the PNS,
i.e.  $r_0 = 31\,\km$, as it is the most favourable place for the
development of the MRI.  At this location, the differential rotation
gradient and the Reynolds numbers are larger than deep inside the PNS.

In the current publication, we discussed some of the simulations
already presented in \citet{Rembiasz_et_al}, but this time with the
initial conditions rescaled to the properties likely present at the
surface of the PNS.  The key physical quantities have been rescaled in
the following way:
\begin{align}
  \label{eq:scalings}
  \MMM &= \tilde{\mathcal{M}}_{r\phi} 
          \left( \frac{r_0}{\tilde{r}_0} \right)^2 
          \left( \frac{\Omega_0}{\tilde{\Omega}_0} \right)^2  
          \left( \frac{\rho_0}{\tilde{\rho}_0} \right)
\\
  b   &= \tilde{ b } \left( \frac{r_0}{\tilde{r}_0} \right)  
                   \left( \frac{\Omega_0}{\tilde{\Omega}_0} \right) 
                   \left( \frac{\rho_0}{\tilde{\rho}_0} \right)^{1/2}
\\
  t   &= \tilde{ t }  \left(  \frac{\Omega_0}{\tilde{\Omega}_0} \right)^{-1}
\\
  \lambdamri &= \tilde{\lambda}_{\mathsmaller{\mathrm{MRI}}} 
                \left( \frac{r_0}{\tilde{r}_0} \right)
\\
  \gammamri  & = \tilde{ \gamma}_{\mathsmaller{\mathrm{MRI}}}  
                \left(  \frac{\Omega_0}{\tilde{\Omega}_0} \right),
\\
L_i &= \bar{L}_i\left( \frac{r}{r_0} \right), 
\end{align}
where quantities with tilde are the values used by
\citet{Rembiasz_et_al}, i.e.  $\tilde{\Omega}_0 = 1824\,\s^{-1}$ and
$\tilde{\rho}_0 = 2.47 \times 10^{13}\,\gccm$, and in the current
paper, we set $\Omega_0 = 767\,\s^{-1}$, and
$\rho_0 = 2.47 \times 10^{12}\,\gccm$.  In \tabref{tab:conversion}, we
present the list of the simulations that were presented in both
papers.
\begin{table*}
  \caption[]{List of the simulations done with \textsc{Aenus} presented in 
    \citet{Rembiasz_et_al} and in the current paper.
    The simulation identifier ,\#, used by \citet{Rembiasz_et_al} 
    and in the current paper, is shown in the left and in the right
    column, respectively. 
  }
%
\begin{center}
\begin{tabular}{|c|c}
\hline
 \citet{Rembiasz_et_al}& current paper 
\\  \hline
5 & A7 \\  
7 & A8a \\  
9 & A9a \\  
10 & A10 \\
11 & A11 \\
\hline
  \end{tabular}
\label{tab:conversion}
\end{center}
 \end{table*}

\section{Technical details}
\label{app:aenus}

\vlc{In this appendix, we assess the quality of the numerical} results
obtained with \textsc{Aenus}.
\vlc{To this end, we analyse in more detail simulation \#A8a}
performed with \textsc{Aenus} (presented \secref{sususeAe}) as well
as, we compare it with simulation \#S19 performed \vlc{with
  \textsc{Snoopy} (\secref{sec:sususeSn}).}

\vlc{In both simulations the MRI channel modes experience a very
  similar evolution with the differences being visible only shortly
  before their termination (see Figs.\,\ref{fig:parasitic} and
  \ref{fig:parasitic_Snoopy}).  The time evolution of the parasitic
  modes} \dvlc{differs} \vlc{almost from the beginning of the
  simulations, however. To investigate this difference in more
  detail,} we note that the dominant axisymmetric MRI modes (with a
vanishing radial component of the wavevector) do not amplify the $v_z$
component of the velocity.
Hence, we could tentatively use this component as a tracer of the
parasitic instabilities if no other (numerical) effects drive the
growth of $v_z$. If we define
\begin{equation}
  \label{eq:11}
  e^{\rm kin}_z\equiv \frac{ \int \frac{1}{2}\rho v_z^2  \der V}{ \int \der V},
\end{equation}
then
\begin{equation}
  \label{eq:10}
  \gamma_z \equiv \frac{ \dot{e}^{\rm kin}_z} { 2 e^{\rm kin}_z} ,
\end{equation}
should in principle be equal to the growth rate of the parasitic
instabilities (provided there are no other effects affecting $v_z$).
\begin{figure}
\centering
\includegraphics[width=1\linewidth]{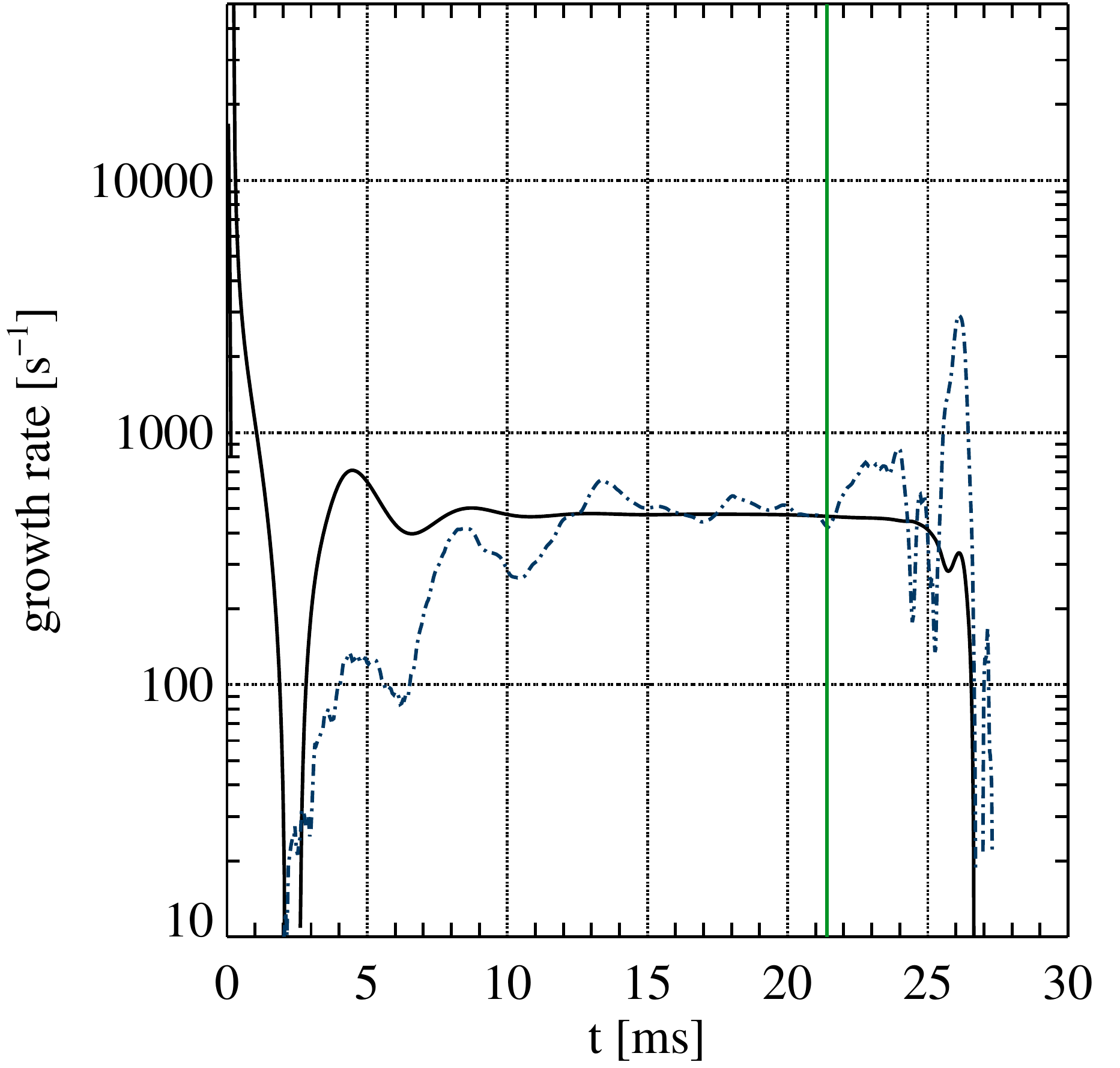} \\
\includegraphics[width=1\linewidth]{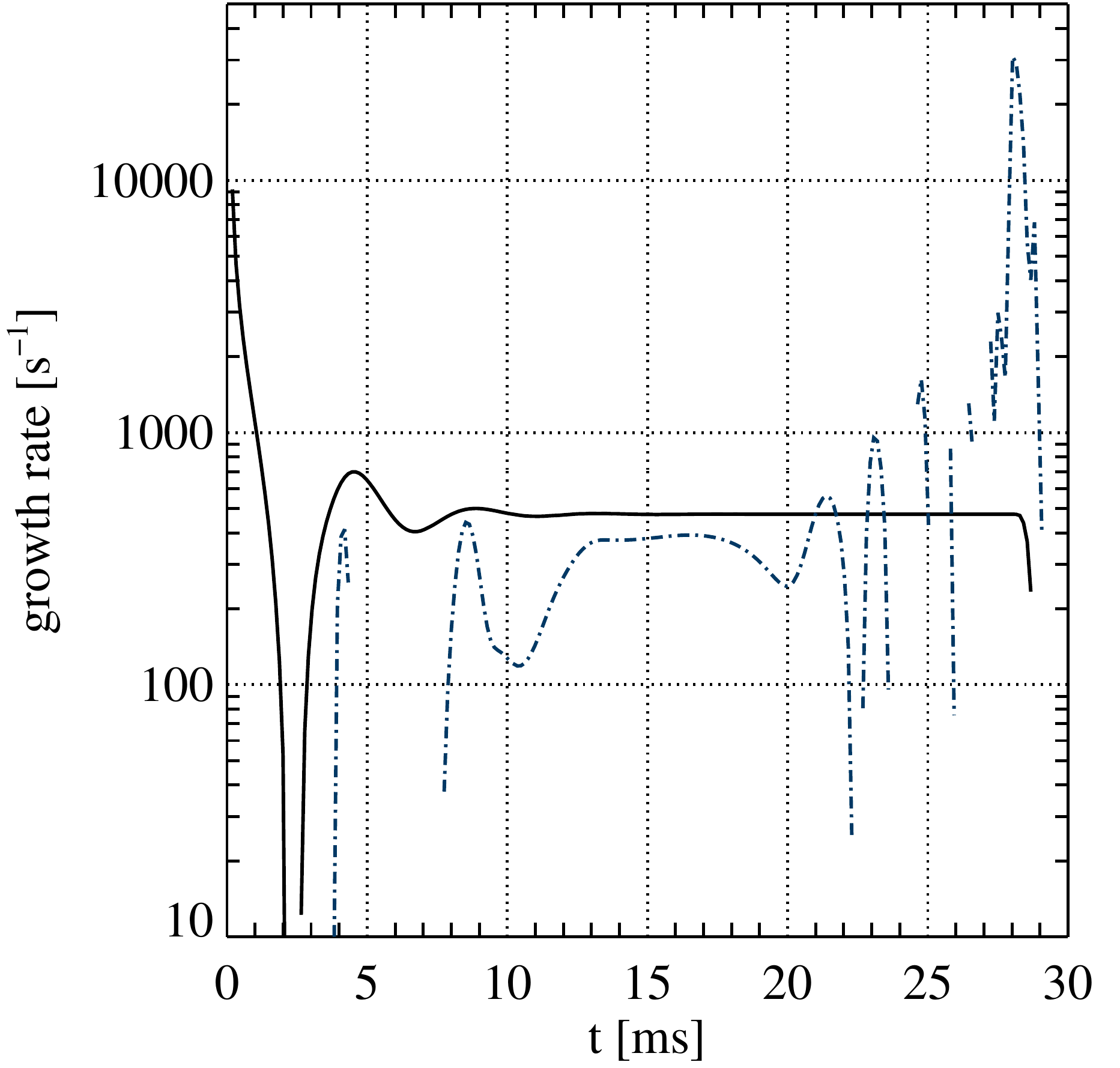} 
\caption{\emph{Top:} Growth rate of the MRI (black line) and growth
  rate of the kinetic energy density in the vertical direction
  (dash-dotted line; Eq.\,\ref{eq:10}) in \textsc{Aenus} simulation
  \#A8a.  Note that to properly compare both growth rates, the MRI
  growth rate is computed from the time evolution of the
  volume-averaged Maxwell stress (Eq.\, \ref{eq:MMM}), instead of from
  the corresponding Fourier modes (as done in \figref{fig:parasitic}).
  The vertical green line marks the time at which the bottom snapshot
  of \figref{fig:04} is taken.  \emph{Bottom:} Same as the upper panel
  but for the \textsc{Snoopy} simulation \#S19.}
\label{fig:white_wave}
\end{figure}
In the upper panel of \figref{fig:white_wave} we provide a comparison
between the time evolution of $\gamma_z$ and that of the MRI growth
rate computed from the box-integrated $v_z$ component of the velocity
field \vlc{in simulation \#A8a}.\footnote{Note the difference with the
  value of the MRI growth rate shown in \figref{fig:parasitic}, which
  is computed from the Fourier transformed $v_z$ component.} It is
evident that $\gamma_z$ follows quite closely the evolution of
$\gammamri$, specially in the time interval between
$t \approx 12\,\ms$ and $t \approx 21\,\ms$, instead of tracing the
evolution of the growth rate of the parasitic instabilities (compare
\figref{fig:white_wave} upper panel with the lower right panel of
\figref{fig:parasitic}). Hence, the growth of $e^{\rm kin}_z$ until
$t \approx 21\,\ms$, cannot be caused (only) by parasitic
instabilities.  Indeed, \citet{Rembiasz} observed a very similar
behaviour of $ e^{\rm kin}_z$ in his 2D and 3D MRI simulations
 \vlc{performed with \textsc{Aenus}} (see
Figs.\, 4.13 and 4.14 therein), which hints that $ e^{\rm kin}_z$ is
amplified by an axisymmetric phenomenon (different from the dominant
MRI axisymmetric modes).

This growth of $ e^{\rm kin}_z$ can be explained by the following
reasons.  Firstly, as MRI magnetic channels grow, they will generate a
non-uniform magnetic pressure accelerating fluid in the vertical
direction (towards magnetic null surfaces of the MRI channels, cf.\
GX94).  Secondly, the initially purely radial profile of the gas
pressure will acquire a vertical gradient due to the advective
transport of internal energy in the up- and down-flows of the channel
modes.
This in turn will create pressure gradients in the vertical direction
and hence vertical fluid motions to redistribute the pressure and
equilibrate the system.
Thirdly, this vertical motions could be related to the radial boundary
conditions.  The MRI growth rate is not constant in the whole
computational domain, as $\gammamri \propto \Omega(r)$.  This means
that the amplitude of the channel modes will grow at a higher rate at
$r = 30\,\km$ than at $r= 32\,\km$.  Consequently, once the MRI
channels are formed, there is a shear at the radial boundaries where
we use periodic boundary conditions for the perturbations.  This shear
introduces some perturbations to all velocity and magnetic field
components, whose influence can be best observed in the $v_z$ velocity
component as it should not be affected by the dominant MRI mode.
During the exponential growth phase, the amplitude of $v_z$ is
proportional to the amplitude of the MRI channels, $\vc$, so that,
$|v_z| \approx 0.15 \vc$.

A comparison of the upper panel of \figref{fig:white_wave} with an
analogous plot for simulation \#S19 \cross{ (discussed in more detail
  in Sec... )} 
done with \textsc{Snoopy} (bottom panel of \figref{fig:white_wave})
provides another argument supporting our hypothesis that those
axisymmetric vertical motions are mostly of the numerical origin. In
the \textsc{Snoopy} simulation, the growth of the vertical motions,
quantitatively indicated by $\gamma_z$, from
$t \approx 12\,\ms \tto 17\,\ms$ is triggered by a subdominant
axisymmetric MRI mode with a non-zero radial component.  Note that
$\gamma_z$ becomes comparable to $\gammamri$ only when the genuine
parasitic instabilities start to grow superexponentially.

However, as much as part of those vertical motions in the
\textsc{Aenus} simulation is clearly of a numerical origin, we should
not be concerned too much with that.  Even though these axisymmetric
modes are dominant 'non-MRI' modes before real parasitic instabilities
appear, they do not cause the MRI termination. In spite of the fact
that the amplitude of these artificial modes is some $15\%$ of the
channel modes, the MRI can grow unaffected by them.

\begin{figure}
\centering
\includegraphics[width=1\linewidth]{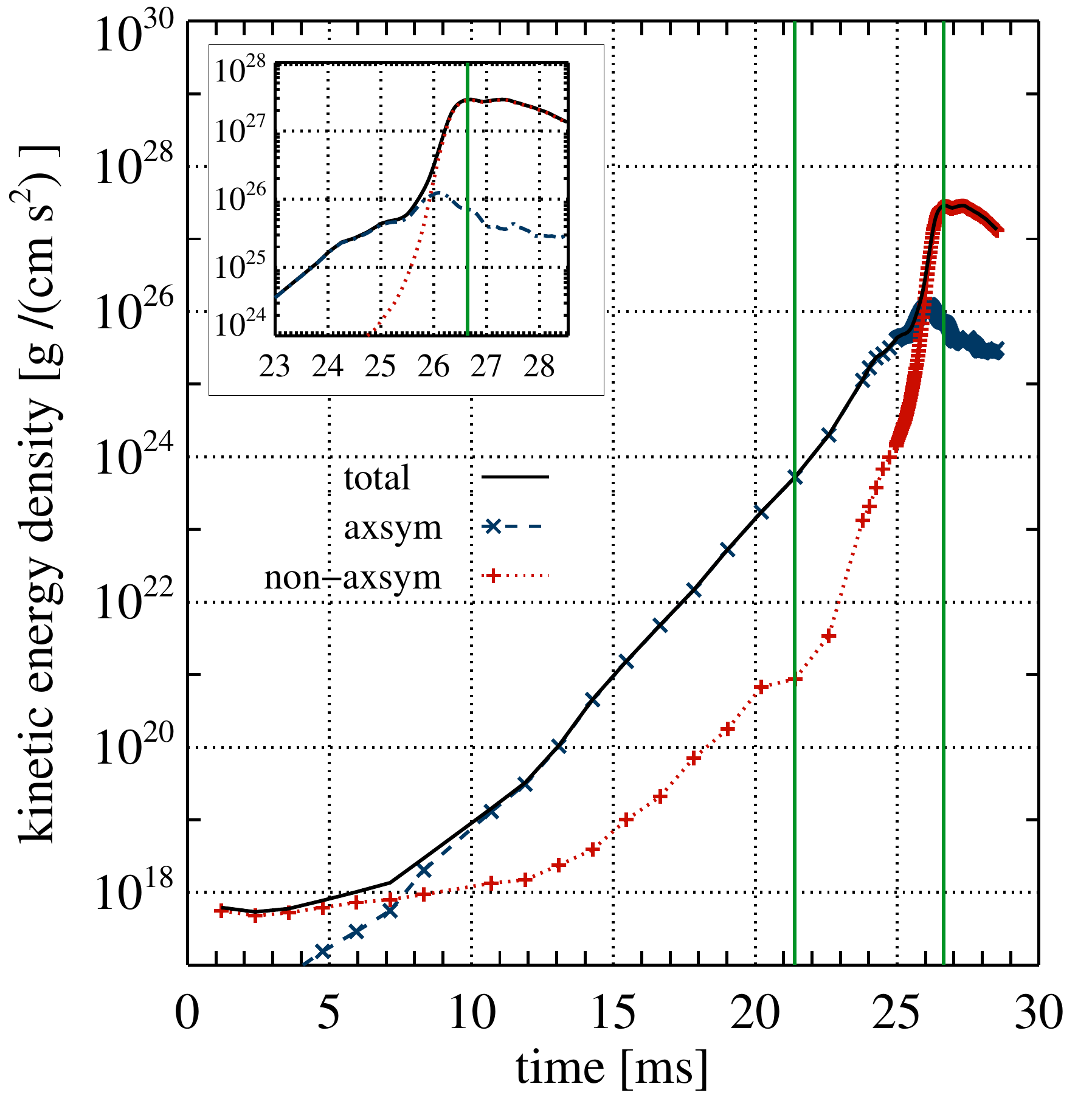} \\
\includegraphics[width=1\linewidth]{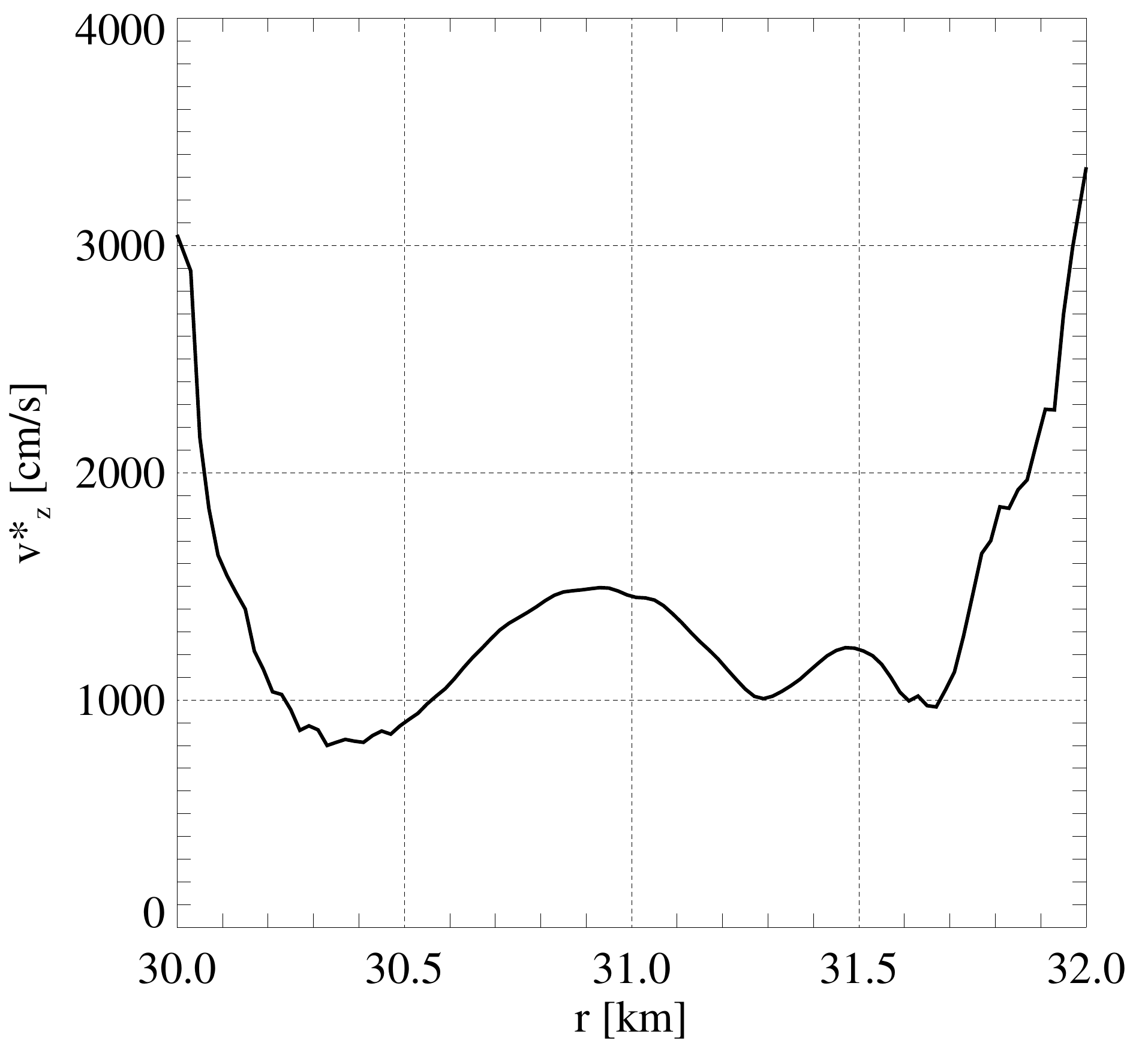} 
\caption{
Top: Evolution of the kinetic energy density, and its axisymmetric
  and non-axisymmetric components for the \textsc{Aenus} model \#A8a.
Bottom: 
Vertically and
  azimuthally averaged RMS amplitude of the non-axisymmetric part
of the velocity component, $\langle v_z(r) \rangle$
  (Eq.~\ref{eq:<v>}) for the same \textsc{Aenus} model as in the upper
  panel.
}
\label{fig:04}
\end{figure}
This can be clearly again seen with the help of Fourier analysis (see
the upper panel of \figref{fig:04}) . Until $t = 23.8\,\ms$, the
axisymmetric modes of $v_z$ account for $99\%$ of the kinetic energy
stored in the $v_z$ component.  However, once the parasitic
instabilities appear and start to contribute to this component, the
fraction of the axisymmetric modes $v_z$ drops significantly and very
rapidly.  At $ t = 25.6\,\ms$, the contribution of axisymmetric modes
to $ e^{\rm kin}_z$ is $86\%$, and only half a millisecond later,
i.e.\ at $ t = 26.2\,\ms$, $16\%$.  At the MRI termination,
i.e.\,$ t = 26.7\,\ms$, the total contribution of the axisymmetric
modes is only $2.5\%$.  This clearly demonstrates that those modes do
not play a decisive role in the MRI termination and that the
instability is terminated by genuine non-axisymmetric parasitic modes.

Another difference between the \textsc{Aenus} and \textsc{Snoopy}
simulations is that in the former, non-axisymmetric $\hat{b}_{\alpha}$
and $\hat{w}_{\alpha}$ modes grow from the beginning of the
simulation, thought until $t\approx 21 \, \ms$ at a rate lower than
$\gammamri$ (upper panels of \figref{fig:parasitic}), whereas in the
latter, these modes basically do not grow until $t\approx 25 \, \ms$
(upper panels of \figref{fig:parasitic_Snoopy}).  This suggest that in
the \textsc{Aenus} simulation, radial boundary conditions introduce
also some non-axisymmetric perturbations (which in turn could be used
as seed perturbations for genuine non-axisymmetric parasitic modes).
To test this hypothesis, we compute the vertically and azimuthally
averaged RMS amplitude of the non-axisymmetric part of the velocity
component $v_z$, i.e.\
\begin{equation}
  \label{eq:<v>}
  \langle v_z(r) \rangle \equiv  \sqrt{ \frac{  \int \int  [
      v_z(r,\phi,z) - \bar{v}_z(r,z) ]^2 \der  \phi 
    \der z  }{ \int \der \phi \int \der z} },
\end{equation}
where 
\begin{equation}
  \bar{ v}_z(r,z) \equiv  \frac{ \int  v_z(r,\phi,z) \der \phi  }{ \int \der \phi },
\end{equation}
before the termination (bottom panel of \figref{fig:04}).  From this
panel, we can see that the non-axisymmetric part of the velocity
component $v_z$ is highest close to the radial boundaries. This
suggest that radial boundary conditions are (most likely) responsible
for triggering this component.  However, only when the growth rate of
the genuine parasitic modes is high enough, the parasitic modes can
use those artificial modes as their initial perturbations.  At later
stages (and at the termination), the non-axisymmetric modes are not
localised at the radial boundaries.

\section*{Acknowledgments}
TR acknowledges support from The International Max Planck Research
School on Astrophysics at the Ludwig Maximilians University Munich,
JG, EM \& TR acknowledge support from the Max-Planck-Princeton Center
for Plasma Physics, and MA, PCD, TR and MO acknowledge support from
the European Research Council (grant CAMAP-259276). We also
acknowledge support from grants AYA2013-40979-P, AYA2015-66899-C2-1-P and
PROMETEOII/2014-069. The authors thank M.~Pessah, C.~McNally and
H.~Latter for helpful discussions. The computations have been
performed at the Leibniz Supercomputing Center of the Bavarian Academy
of Sciences and Humanities (LRZ), the Rechenzentrum Garching of the
Max-Planck-Gesellschaft (RZG), and at the Servei d'Inform\`atica of
the University of Valencia.


\bibliographystyle{mn2e}

\label{lastpage}

\end{document}